\documentclass[onecolumn,11pt,draftcls,journal]{IEEEtran}
 \usepackage{ifpdf}
    \ifpdf
    \else

\usepackage{array}
\usepackage{cite}
\usepackage{amsfonts}
\usepackage{amsmath}
\usepackage{amssymb}
\usepackage{amsxtra}
\usepackage{varioref}
\usepackage{multicol}
\usepackage{float}
\usepackage{rotate}
\usepackage{color}
\usepackage{enumerate}
\usepackage[active]{srcltx} 
\usepackage{mathrsfs}
\usepackage{color}
\usepackage{rotating}
\usepackage{Milancommands}

\usepackage{times}
\usepackage[english]{babel}
\usepackage{inputenc}
\usepackage{fontenc}

\newtheorem{thm}{Theorem}
\newtheorem{defn}{Definition}

\newtheorem{lem}{Lemma}
\newtheorem{rem}{Remark}
\newtheorem{prop}{Proposition}

\newtheorem{opprob}{\textbf{Optimization Problem}}

\title{Improved Upper Bounds to the Causal Quadratic Rate-Distortion Function for Gaussian Stationary Sources }

\author{Milan S. Derpich%
\thanks{Milan S. Derpich is with the Department of Electronic Engineering,
Universidad T\'ecnica Federico Santa Mar\'ia, Casilla 110-V, Valpara\'iso, Chile 
(email: milan.derpich@usm.cl).  } 
and Jan {\O}stergaard%
\thanks{Jan {\O}stergaard is with the Department of Electronic Systems, Aalborg University,
Aalborg, Denmark
(email: janoe@ieee.org)
This research was partially supported  by FONDECYT project 3100109, Anillo project ACT-53/2010,
and by
the Danish Research Council for Technology and Production Sciences, grant no. 274-07-0383.
}
}%

\begin{document}
	\maketitle
\begin{abstract}
We improve the existing achievable rate regions for causal and for zero-delay source coding of stationary Gaussian sources under an average mean squared error (MSE) distortion measure.
To begin with, we find a closed-form expression for the information-theoretic causal rate-distortion function (RDF) under such distortion measure, denoted by $R_{c}^{it}(D)$, for first-order Gauss-Markov processes.
$R_{c}^{it}(D)$ is a lower bound to the optimal performance theoretically attainable (OPTA) by any causal source code, namely $R_{c}^{op}(D)$.
We show that, for Gaussian sources, the latter can also be upper bounded as $R_{c}^{op}(D) \leq R_{c}^{it}(D) + 0.5\log_{2}(2\pi \expo{})$~bits/sample.
In order to analyze $R_{c}^{it}(D)$ for arbitrary zero-mean Gaussian stationary sources,
we introduce $\overline{R_c^{it}}(D)$, the information-theoretic causal RDF when the reconstruction error is jointly
stationary with the source.
Based upon $\overline{R_c^{it}}(D)$, 
we derive three closed-form upper bounds to the additive rate loss defined as $\overline{{R}_c^{it}}(D) - R(D)$, where $R(D)$ denotes Shannon's RDF.
Two of these bounds are strictly smaller than $0.5$ bits/sample at all rates.
These bounds differ from one another in their tightness and ease of evaluation;
the tighter the bound, the more involved its evaluation.
We then show that, for any source spectral density and 
any positive distortion $D\leq \sigma_{\rvax}^{2}$,
 $\overline{R_c^{it}}(D)$ can be realized by an AWGN channel surrounded by a unique set of causal \mbox{pre-,} \mbox{post-,} and feedback filters.
We show that finding such filters constitutes a convex optimization problem. 
In order to solve the latter, we propose an iterative optimization procedure that yields the optimal filters and is guaranteed to converge to $\overline{R_c^{it}}(D)$.
Finally, by establishing a connection to feedback quantization we 
design a causal and a zero-delay coding scheme which, for Gaussian sources, achieves an operational rate lower than 
$\overline{R_c^{it}}(D) + 0.254$ and  $\overline{R_c^{it}}(D) + 0.754$ bits/sample, respectively.
This implies that the OPTA among all \emph{zero-delay} source codes, denoted by $R_{zd}^{op}(D)$, is upper bounded as $R_{zd}^{op}(D)< \overline{R_c^{it}}(D) + 1.254 < R(D) + 1.754$ bits/sample.
\end{abstract}


\section{Introduction}\label{sec:intro}
In zero-delay source coding, the reconstruction of each input sample must take place at the same time instant the corresponding input sample has been encoded.
Zero-delay source coding is
desirable in many applications, e.g., in real-time applications where one cannot afford to have large
delays~\cite{huajac04}, or in systems involving feedback, in which the current input depends on the previous
outputs~\cite{tatiko00,tatsah04,silder11}.
A weaker notion closely related to the principle behind zero-delay codes is that of causal source coding, wherein the reproduction of the present source sample \emph{depends} only on the present and past source samples but not on the future source samples~\cite{neugil82,linzam06}.
This notion does not preclude the use of non-causal entropy coding, and thus it does not guarantee zero-delay reconstruction.
Nevertheless, any zero-delay source code must also be causal.

It is known that, in general, causal codes cannot achieve the \emph{rate-distortion function} (RDF) $R(D)$ of the source, which is the \emph{optimal  performance theoretically attainable} (OPTA) in the absence of causality constraints~\cite{berger71}.
However, it is in general not known how close to $R(D)$ one can get when restricting attention to the class of causal or
zero-delay source codes, except, for causal codes, when dealing with
memory-less sources~\cite{neugil82}, stationary sources at high resolution~\cite{linzam06}, or first-order Gauss-Markov
sources under a per-sample MSE distortion metric~\cite{tatsah04}.

For the case of memory-less sources, it was shown by Neuhoff and Gilbert that the optimum rate-distortion performance of
causal source codes is achieved by time-sharing at most two memory-less scalar quantizers (followed by entropy
coders)~\cite{neugil82}.
In this case, the rate loss due to causality was shown to be given by the space-filling loss of the quantizers, i.e. the
loss is at most $(1/2)\ln(2 \pi e/12)$ ($\simeq$ 0.254) bits/sample.
For the case of Gaussian stationary sources with memory and MSE distortion, Gorbunov and Pinsker showed that the
information theoretic%
\footnote{Here and in the sequel, the term ``information theoretic'' refers to the use of 
mutual information as a measure of the rate.}
causal RDF, here denoted by $R_{c}^{it}(D)$ and to be defined formally in
Section~\ref{sec:prelimi}, tends to Shannon's RDF as the distortion goes to
zero~\cite{pingor87,gorpin91}.
The possible gap between the OPTA of causal source codes and this information-theoretic causal RDF was not assessed.
On the other hand, for arbitrary stationary sources with
finite differential entropy and under high-resolution
conditions, it was shown in~\cite{linzam06} that the
rate-loss of causal codes (i.e, the difference between their
OPTA and Shannon's RDF) is at most the space-filling loss of
a uniform scalar quantizer.
With the exception of memory-less sources and first-order Gauss-Markov sources,
the ``price'' of causality at general rate regimes for other stationary sources remains an open problem.
 However, it is known that for any source, the mutual information rates across an \textit{additive white Gaussian noise} (AWGN) channel 
 and across a scalar ECDQ channel do not exceed $R(D)$ by more than $0.5$ and $0.754$ bits per sample, respectively~\cite{zamkoc08},~\cite{zamfed92}. 
This immediately yields the bounds $R_{c}^{it}(D)\leq R(D)+0.5$ and $R_{c}^{op}(D)\leq R(D)+0.754$.

In causal source coding it is generally difficult to provide a constructive proof of achievability since Shannon's      
random codebook construction, which relies upon jointly encoding long sequences
of source symbols, is not directly applicable even in the case of memory-less sources.
Thus, even if one could obtain an outer bound for the achievable region based on an information theoretic RDF,
finding the inner bound, i.e., the OPTA, would still remain a challenge. 

There exist other results related to the information-theoretic causal RDF, in which achievability is not addressed.
The minimum sum rate 
necessary to sequentially encode and decode two
scalar correlated random variables under a coupled fidelity criterion was studied in~\cite{visber00}.
A closed-form expression for this minimum rate is given in~\cite[Theorem~4]{visber00} for the special case of a squared
error distortion measure and a per-variable (as opposed to a sum or average) distortion constraint.
In~\cite{tatiko00}, the minimum rate for causally encoding and decoding source samples
(under per-sample or average distortion constraints) was given the
name~\emph{sequential rate-distortion function} (SRDF).
Under a per-sample MSE distortion constraint $D$,
it was also shown in~\cite[p.~187]{tatiko00} that
for a first-order Gauss-Markov source $\rvax(k+1) = a_{1}\rvax(k) + \xi(k)$, where 
$\set{\xi(k)}$
is a zero-mean white Gaussian process with variance $\sigsq_{\xi}$, the information theoretic SRDF%
\footnote{The information theoretic SRDF is the one defined in~\cite[Def.~5.3.1]{tatiko00}, where
it is denoted by $R^{SRD}_{T,N}(D)$.}
$R_{SRD}^{it}(D)$  takes the form
\begin{align}\label{eq:Rsrd}
 R_{SRD}^{it}(D) = \min\Set{0\;,\;\frac{1}{2}\log_{2}\left(a_{1}^{2} +\frac{\sigsq_{\xi}}{D}
\right)}
\fspace \textrm{bits/sample}
,
\end{align}
for all $D>0$.%
\footnote{
It has not been established whether~\eqref{eq:Rsrd} is achievable or how close one can get to it.}
No expressions are known for $R_{SRD}^{it}(D)$ for higher-order Gauss-Markov sources.
Also, with the exception of memory-less Gaussian sources, $R_{c}^{it}(D)$, with its average MSE distortion constraint (weaker
than a per-sample MSE constraint), has not been characterized.

In this paper, we improve the existing inner and outer rate-distortion bounds for causal and for zero-delay source coding of zero-mean Gaussian stationary sources and average MSE distortion. 
We start by 
showing that, 
for any zero-mean Gaussian source with bounded differential entropy rate, the causal OPTA exceeds $R_{c}^{it}(D)$ by less than approximately $0.254$ bits/sample.
Then we revisit the SRDF problem for first-order Gauss-Markov sources under a per-sample distortion constraint schedule and find the explicit expression for the corresponding RDF by means of an alternative, 
constructive derivation. 
This expression, which turns out to differ from the one found in~\cite[bottom of~p.~186]{tatiko00}, 
allows us to show that  
for first-order Gauss-Markov sources, the information-theoretic causal RDF
$R_{c}^{it}(D)$ for an average (as opposed to per-sample) distortion measure coincides
with~\eqref{eq:Rsrd}.
In order to upper bound $R_{c}^{it}(D)$ for general Gaussian stationary sources,
we introduce the
information-theoretic causal RDF when the distortion is jointly stationary with the source and
denote it by $\overline{R_c^{it}}(D)$.
We then derive three closed-form upper bounding functions to the rate-loss
$\overline{R_c^{it}}(D)-R(D)$, which can be applied to any stationary Gaussian random process.
Two of these bounds are, at all rates, strictly tighter than the best
previously known general bound of $0.5$ bits/sample.
Since, by definition, $R_{c}^{it}(D)\leq \overline{R_{c}^{it}}(D)$, we have that  
\begin{align}\label{Antar}
 R_{c}^{it}(D)-R(D) 
\overset{(a)}{\leq}  \overline{R_c^{it}}(D) - R(D),
\end{align}
and thus all four three bounding functions also upper bound the gap
$R_{c}^{it}(D)-R(D)$.
As we shall see,
equality holds in $(a)$ if $R_{c}^{it}(D)$ could be realized by a test channel with distortion jointly stationary with the source, which seems a reasonable conjecture for stationary sources.

We do not provide a closed-form expression for
$\overline{R_c^{it}}(D)$ (except for first-order Gauss-Markov
sources), and thus the upper bound on the \emph{right-hand-side} (RHS)
of~\eqref{Antar}
(the tightest bound discussed in this paper) is
not evaluated analytically for the general case.
However, we propose an iterative procedure that can be implemented
numerically and which allows one to evaluate
$\overline{R_c^{it}}(D)$, for any source \emph{power spectral density} (PSD) and $D>0$, with any desired accuracy.
This procedure is based upon the iterative optimization of causal  pre-, post- and feedback-filters around an AWGN channel.
A key result in this paper (and its second main contribution) is showing that such filter optimization problem is convex in the frequency responses of all the filters.
This guarantees that the 
mutual information rate between source and reconstruction
yielded by our iterative procedure converges monotonically to $\overline{R_c^{it}}(D)$ as the number of iterations and the order of the filters tend to infinity.
This equivalence between the solution to
a convex filter design optimization problem and
$\overline{R_{c}^{it}}(D)$ avoids the troublesome
minimization over mutual informations, thus making it possible to
actually compute $\overline{R_{c}^{it}}(D)$ in practice, for general Gaussian stationary sources.
We then make the link between $\overline{R_c^{it}}(D)$ and the OPTA of causal and zero-delay codes.
More precisely, 
when the AWGN channel is replaced by a subtractively dithered uniform scalar quantizer followed by memory-less entropy coding, the filters obtained with the iterative procedure yield a causal source coding system whose operational rate is below $\overline{R_{c}^{it}}(D)+(1/2)\log_{2}(2\pi\expo{})$~bits/sample.
If the entropy coder in this system is restricted to encode quantized values individually (as opposed to long sequences of them), then this system achieves zero-delay operation with an operational rate below $\overline{R_{c}^{it}}(D)+(1/2)\log_{2}(2\pi\expo{})+1$~bits/sample.
This directly translates into an upper bound to the OPTA of zero-delay source codes, namely $R_{zd}^{op}(D)$.
To illustrate our results, we present an example for a zero-mean \mbox{AR-1} and a zero-mean \mbox{AR-2} Gaussian
source, for which we evaluate the closed-form bounds and obtain an approximation of $\overline{R_{c}^{it}}(D)$
numerically by applying the iterative procedure proposed herein.

This paper is organized as follows: In Section~\ref{sec:prelimi}, we review some preliminary notions.
We prove in section~\ref{sec:Upr_bonds_for:Rop_from_Rit} that the OPTA for Gaussian sources does not exceed the information-theoretic RDF by more than approximately 0.254 bits per sample.
Section~\ref{sec:Rcit(D)_1st_order} contains the derivation of a closed-form expression for $R_{c}^{it}(D)$ for first-order Gauss-Markov sources.
In Section~\ref{sec:Analytic_Bounds} we formally introduce $\overline{R_c^{it}}(D)$ and derive the three closed-form upper bounding functions for the information-theoretic rate-loss of causality.
Section~\ref{sec:convexity} presents the iterative procedure to calculate $\overline{R_c^{it}}(D)$, after presenting the proof of convexity that guarantees its convergence.
The two examples are provided in Section~\ref{sec:Example}.
Finally, Section~\ref{sec:concl} draws conclusions.
(Most of the proofs of our results are given in sections~\ref{sec:proof_italwaysbounds} to~\ref{sec:proof_Jsc()_is_convex}.)

\subsection*{Notation}
$\Rl$ and $\Rl_{0}^{+}$ denote, respectively, the set of real numbers and the set of non-negative real numbers.
$\Z$ and $\Z^{+}$ denote, respectively, the sets of integers and positive integers.
We use non-italic lower case letters, such as $\rvax$, to denote scalar random variables, and boldface lower-case and upper-case letters to denote vectors and matrices, respectively.
We use $\bA^{\dagger}$, $\ospn\set{\bA}$ and $\Nsp\set{\bA}$ to denote the Moore-Penrose pseudo-inverse, the column span and the null space of the matrix $\bA$, respectively.
The expectation operator is denoted by $\Expe{\,}$.
The notation $\sigsq_{\rvax}$ refers to the variance of $\rvax$.
The notation $\set{\rvax(k)}_{k=1}^{\infty}$ describes a one-sided random process, which may also be written simply as $\procx$.
We write $\rvax^{k}$ to refer to the sequence $\set{\rvax(i)}_{i=1}^{k}$.
The PSD of a wide-sense stationary process $\procx$ is denoted by $S_{\rvax}\ejw, \, \w\in\pipi$.
Notice that $\sigsq_{\rvax}=\intpipi{S_{\rvax}\ejw}$.
For any two functions $f,g:\pipi\to\mathbb{C}$, $f,g\in\Ldos$, we write the standard squared norm and inner product as
$
\norm{f}^{2}\eq \intpipi{\abs{f(\w)}^{2}}
$
and
$
\ip{f,g}\eq\intpipi{f(\w)g(\w)^{\ast}}$,
respectively, where~$^{\ast}$ denotes complex conjugation.
For one-sided random processes $\procx$ and $\procy$, the term
$\bar{I}(\procx ;\procy)= \lim_{k\to\infty}\sup \frac{1}{k}I(\rvax_{1}^{k} ; \rvay_{1}^{k})$
denotes the mutual information rate between $\procx$ and $\procy$, provided the limit exists.
Similarly, for a stationary random process $\procx$, $\bar{h}(\procx)=\lim_{k\to\infty} h(x(k)|x^{k})$ denotes the differential entropy rate of $\procx$.

\section{Preliminaries}\label{sec:prelimi}
A source \emph{encoder-decoder} (ED) pair encodes a source $\set{\rvax(k)}_{k=-\infty}^{\infty}$ into binary symbols, from which a reconstruction   $\set{\rvay(k)}_{k=1}^{\infty}$ of $\set{\rvax(k)}_{k=1}^{\infty}$ is generated.
The end-to-end effect of any ED pair can be described by a series of \emph{reproduction functions} $\set{f_{k}}_{k=1}^{\infty}$, such that, for every $k\in\Z^{+}$,
\begin{align}\label{eq:repro_func_def}
	\rvay_{1}^{k} = f_{k}(\rvax_{-\infty}^{\infty}),
\end{align}
where we write $\rvay_{i}^{k}$ as a short notation for $\set{\rvay(j)}_{j=i}^{k}$.
Following~\cite{neugil82}, we say that an ED pair is \emph{causal} if and only if it satisfies the following
definition~\cite{neugil82}:
\begin{defn}[Causal Source Coder]\label{def:Causal_SC}
An ED pair is said to be causal if and only if its reproduction functions are such that
\begin{align*}
	f_{k}(\rvax_{-\infty}^{\infty}) = f_{k}(\tilde{\rvax}_{-\infty}^{\infty}), \fspace \textrm{ whenever } \rvax_{-\infty}^{k}
    =
    \tilde{\rvax}_{-\infty}^{k}, \fspace \forall k\in\Z^{+}.
\end{align*}
\finenunciado
\end{defn}
It also follows from Definition~\ref{def:Causal_SC} that an ED pair is causal if and only if the following Markov chain
holds for every possible random input process $\procx$:
\begin{align}\label{eq:Markov_Causal}
	\rvax_{k+1}^{\infty}\to \rvax^{k}_{-\infty}\to \rvay_{1}^{k}, \fspace \forall k\in\Z^{+}.
\end{align}
It is worth noting that if the reproducing functions are random, then this equivalent causality constraint must require that~\eqref{eq:Markov_Causal} is satisfied for each realization of the reproducing functions $\set{f_{k}}_{k=1}^{\infty}$.

Let~%
 $\rvaL_{k}(\rvax_{1}^{\infty})$ be the total number of bits that the decoder has received when it generates the output subsequence $\rvay_{1}^{k}$.
Define  $\rveb(k)\in\set{0,1}^{\rvaL_{k}}$ as the random binary sequence that contains the bits that the decoder has received when $\rvay_{1}^{k}$ is generated.
Notice that $\rvaL_{k}$ is, in general, a function of all source samples, since the binary coding may be non-causal, i.e., $\rvay_{1}^{k}$ may be generated only after the decoder has received enough bits to reproduce $\rvay_{1}^{m}$, with $m > k$.
We highlight the fact that even though $\rveb(k)$ may contain bits which depend on samples 
$\rvax(\ell)$ with $\ell> k$, the random sequences $\rvax_{-\infty}^{\infty}$ and $\rvay_{1}^{k}$ may still satisfy~\eqref{eq:Markov_Causal}, i.e., the ED pair can still be causal.
Notice also that $\rvaL_{k}(\rvax_{1}^{\infty})$ is a random variable, which depends on $\rvax_{-\infty}^{\infty}$, the functions $\set{f_{k}}$ and on the manner in which the source is encoded into the binary sequence sent to the decoder.

For further analysis, we define the \emph{average operational rate} of an ED pair as~\cite{neugil82}
\begin{align}\label{eq:r_def}
	r(\procx,\procy ) \eq \lim_{k\to\infty} \sup \frac{1}{k}\Expe{L_{k}(\rvax_{-\infty}^{\infty})  }.
\end{align}
In the sequel, we focus only on the MSE as the distortion measure.
Accordingly, we define the \emph{average distortion} associated with an ED pair as:
\begin{align}\label{eq:d_def}
	d(\procx,\procy ) \eq \lim_{k\to\infty}\sup \frac{1}{k}\Expe{ \norm{\rvax_{1}^{k} - \rvay_{1}^{k} }^{2}}.
\end{align}
The above notions allow us to define the operational causal RDF as follows:
\begin{defn}
The Operational Causal Rate-Distortion Function for a source $\procx$ is defined as~\cite{neugil82}:
\begin{align}\label{eq:Rc_op_def}
	R_{c}^{op}(D) 
	\eq 
	\inf_{
	\substack{
	\procy : \rvay(k)=f_{k}(\rvax^{k}),\forall k\in \Z^{+}\\
	\set{f_{k}} \textrm{ causal} ,\\
	d(\procx,\procy ) \leq D.
	}
	}r(\procy,\procx ).
\end{align}
\finenunciado
\end{defn}
We note that the operational causal rate distortion function defined above corresponds to the OPTA of all causal ED pairs.

In order to find a meaningful information-theoretical counterpart of $R_{c}^{op}(D)$, we note
from~\cite[Theorem~5.3.1]{covtho06} that
\begin{align}\label{eq:coding_ineq}
	\frac{1}{k}\Expe{L_{k}(\rvax_{1}^{\infty})} \geq \frac{1}{k}H(\rveb(k)), \fspace \forall k\in\Z^{+}.
\end{align}
Also, from the Data Processing Inequality~\cite{covtho06}, it follows immediately that
\begin{align}\label{eq:H_and_I}
	H(\rveb(k)) = I(\rveb(k);\rveb(k)) \geq I(\rvax_{1}^{\infty} ; \rvay_{1}^{k}) \geq I(\rvax_{1}^{k} ; \rvay_{1}^{k}), 
\end{align}
where the last inequality turns into equality for a causal ED pair, since in that case~\eqref{eq:Markov_Causal} holds.
Thus, combining~\eqref{eq:r_def},~\eqref{eq:coding_ineq} and~\eqref{eq:H_and_I},
\begin{align}\label{eq:r_>_Ixy}
r(\procx,\procy )
\geq
\lim_{k\to\infty}\sup \frac{1}{k}I(\rvax_{1}^{k} ; \rvay_{1}^{k}) 
=
\bar{I}(\procx;\procy) .
\end{align}
This lower bound motivates the study of an information-theoretic causal rate distortion function, as defined below.
\begin{defn}\label{def:CausalRDF}
\emph{
	The Information-Theoretic Causal Rate-Distortion Function for a source $\procx$, with respect to the average MSE distortion measure, is
	defined as
	\begin{align*}
		R_{c}^{it}(D) \eq \inf\bar{I}(\procx;\procy),
	\end{align*}
where the infimum is over all processes $\procy$ such that $d(\procx,\procy)\leq D$  and such that~\eqref{eq:Markov_Causal} holds.}
\finenunciado
\end{defn}
The above definition is a special case of the 
non-anticipative epsilon-entropy
introduced by Pinsker and Gorbunov, which was shown to converge to Shannon's RDF, for Gaussian stationary sources 
and in the limit as the rate goes to infinity~\cite{pingor87,gorpin91}.

In the non-causal case, it is known
that for any source and for any single-letter distortion measure, the OPTA equals the information-theoretic RDF~\cite{covtho06}.
Unfortunately, 
such a
strong equivalence between the OPTA and the information-theoretic RDF does not seem to be possible in the causal case (i.e., for $R_{c}^{it}(D)$).
(One exception is if one is to jointly and causally encode an asymptotically large number of parallel Gaussian sources, in which case $R_{c}^{it}(D)$  can be shown to coincide with the OPTA of causal codes.)
Nevertheless, as outlined in Section~\ref{sec:intro}, it is possible to obtain lower and upper bounds to the OPTA of causal codes from $R_{c}^{it}(D)$.
Indeed, and to begin with,
since $R_{c}^{it}(D)\geq R(D)$, it follows directly from~\eqref{eq:Rc_op_def} and~\eqref{eq:r_>_Ixy} that
\begin{align}\label{eq:Rcop_>_Rcit}
	R^{op}_{c}(D)\geq R_{c}^{it}(D)\geq R(D).
\end{align}
The last inequality in~\eqref{eq:Rcop_>_Rcit} is strict, in general, and becomes equality when the source is white or when the rate tends to infinity.
Also, as it will be shown in Section~\ref{sec:Upr_bonds_for:Rop_from_Rit}, for Gaussian sources $R_{c}^{op}(D)$ does not exceed $R_{c}^{it}(D)$ by more than approximately $0.254$ bits/sample, and thus an upper bound to $R_{c}^{op}(D)$ can be obtained from $R_{c}^{it}(D)$.

For completeness, and for future reference, we recall that 
for any MSE distortion $D>0$, the RDF for a stationary Gaussian source with PSD $S_{\rvax}\ejw$ is equal to the
associated information-theoretic RDF, given by the ``reverse water-filling'' equations~\cite{berger71}
\begin{subequations}\label{eq:R(D)waterfill}
 \begin{align}
 R(D) & = \frac{1}{4\pi}
 \Intfromto{-\pi}{\pi}
\max\Set{ 0\, ,\, 
\log_{2}\left(\frac{S_{\rvax}\ejw}{\theta} \right)
}
d\w \label{eq:R(D)waterfill_R(D)}
\\
D&=
\Intpipi{\min\Set{\theta\,,\,S_{\rvax}\ejw  }  }.
\end{align}
\end{subequations}

Although in general it is not known
by how much $R_{c}^{it}(D)$ exceeds $R(D)$, 
for Gaussian stationary sources one can readily find an upper bound for $R_{c}^{it}(D)$ in the quadratic Gaussian RDF for source-uncorrelated distortion, defined as~\cite{derost08}
\begin{align}\label{eq:Rperpdef}
 R^{\perp}(D)\eq \inf_{\procy} \bar{I}(\procx,\procy), 
\end{align}
where the infimum is taken over all output processes $\procy$ consistent with MSE$\leq D$ and such that the reconstruction error $\set{\rvay(k)-\rvax(k)}$ is uncorrelated with the source.
More precisely, it is shown in~\cite{derost08} that this RDF, given by 
\begin{subequations}\label{eq:R^{perp}}
 \begin{align}
 R^{\perp}(D) &=\frac{1}{2\pi}\intfromto{-\pi}{\pi}
 \log\left(  \frac{\hksqrt{S_{X}(\w) + \alpha} + \hksqrt{S_{X}(\w)} }
				{\hksqrt{\alpha}} \right)d\w,
\end{align}
wherein $\alpha>0$ is the only scalar that satisfies
\begin{align}
 D & 
= \frac{1}{4\pi}\intfromto{-\pi}{\pi}
{\left(\hksqrt{S_{X}(\w) +\alpha } - \hksqrt{S_{X}(\w)} \right)\hksqrt{S_{X}(\w)}                 }d\w,
\end{align}
\end{subequations}
can be realized causally.

More generally, it is known that, for any source, the mutual information across an AWGN channel (which satisfies~\eqref{eq:Markov_Causal}) introducing noise with variance~$D$, say $R_{AWGN}(D)$, exceeds Shannon's RDF $R(D)$ by at most $0.5$ bits/sample, see, e.g.~\cite{zamfed92}.
Thus, we have:
\begin{align}\label{eq:bounds_Rop_Rit_RD}
	R_{c}^{it}(D)\leq R_{AWGN}(D) \leq R(D) + 0.5 \fspace \textrm{bits/sample}, \fspace\forall D>0.
\end{align}
Until now it has been an open question whether a bound tighter than~\eqref{eq:bounds_Rop_Rit_RD}  can be obtained for sources with memory and at general rate regimes~\cite{zamkoc08}.
In sections~\ref{sec:Rcit(D)_1st_order}, \ref{sec:Analytic_Bounds} and~\ref{sec:convexity},
we show that for for Gaussian sources this is indeed the case.
But before focusing on upper bounds for $R_{c}^{it}(D)$, its operational importance will be established by showing in the following section that, for Gaussian sources, the OPTA does not exceed $R_{c}^{it}(D)$ by more than approximately $0.254$ bits/sample.

\section{Upper Bounds to $R_{c}^{op}$ from $R_{c}^{it}$}\label{sec:Upr_bonds_for:Rop_from_Rit}
 In this section we show that, for any Gaussian source $\procx$ and $D\geq 0$, an upper bound to $R_{c}^{op}$ can be readily obtained from $R_{c}^{it}(D)$ by adding (approximately) $0.254$ bits per sample to $R_{c}^{it}(D)$.
This result is first formally stated and proved for finite subsequences of any Gaussian source.
Then, it is extended to Gaussian stationary processes.

We start with two definitions.
\begin{defn}
 The \textit{causal information theoretic RDF} for a zero-mean Gaussian random vector of length $\ell$ is defined as
\begin{align}
     R_{c}^{it(\ell)}(D) &= \inf \tfrac{1}{\ell}I(\rvex;\rvey),
\end{align}
where the infimum is taken over all output vectors satisfying
the causality constraint
\begin{align}\label{eq:MCcausal_ell}
\rvay(k)
\leftrightarrow
\rvax^{k}
\leftrightarrow
\rvax_{k+1}^{\ell}
,\;\forall k=1,\ldots,\ell-1
\end{align}
and the distortion constraint
\begin{align}\label{eq:dconstrell}
 d(\rvex,\rvey)\eq \frac{1}{\ell}\Expe{\norm{\rvey-\rvex}^{2}}\leq D.
\end{align}
\finenunciado
\end{defn}

\begin{defn}
The \textit{operational causal RDF} for a zero-mean Gaussian random vector of length $\ell$ is defined as
\begin{align}
     R_{c}^{op(\ell)}(D) &= 
     \inf_{
     \substack{
	\rvay_{1}^{k} : \rvay(k)=f_{k}(\rvax^{k}),\forall k=1,\ldots,\ell\\
	\set{f_{k}} \textrm{ causal} ,\\
	d(\rvex,\rvey ) \leq D.
	}
    } 
     r(\rvax^{k},\rvay^{k})
\end{align}
\finenunciado
\end{defn}

We will also need the following result~\cite[Lemma~1]{derost08}:
\begin{lem}\label{lem:ZisGaussian}
Let $\rvex\in\Rl^{\ell}\sim\Nsp(\bzero,\bK_{\rvex})$. 
Let $\rvez\in\Rl^{\ell}$ and $\rvez_{G}\in\Rl^{\ell}$
be two random vectors with zero mean and the same covariance matrix, i.e.,
$\bK_{\rvez}=\bK_{\rvez_{G}}$, and having the same cross-covariance matrix with respect to $\rvex$,
that is, $\bK_{\rvex,\rvez}=\bK_{\rvex,\rvez_{G}}$.
If $\rvez_{G}$ and $\rvex$ are jointly Gaussian, and if $\rvez$ has any distribution, then
%
\begin{align}
 I(\rvex;\rvex+\rvez) \geq I(\rvex;\rvex+\rvez_{G}).\label{eq:Iineq}
\end{align}
If furthermore  $\abs{\bK_{\rvex+\rvez}}>0$, then equality is achieved in~\eqref{eq:Iineq} if and
only if $\rvez\sim\Nsp(\bzero,\bK_{\rvez})$ with $\rvez$ and $\rvex$ being jointly Gaussian.
\finenunciado
\end{lem}
Notice that if one applies Lemma~\ref{lem:ZisGaussian} to a reconstruction error with which the output sequence
satisfies the causality constraint~\eqref{eq:Markov_Causal}, then the Gaussian version of the same
reconstruction error will also produce an output causally related with the input.
More precisely, if a given reconstruction error $\rvaz^{\ell}$ satisfies~\eqref{eq:Markov_Causal},
then, for all $j\leq k<i\leq N$, it holds that
$
0
=
\Expe{\rvax(i)(\rvay^{j} -\Expe{\rvay^{j}|\rvax^{k}})}
=
\Expe{\rvax(i)(\rvax^{j}+\rvaz^{j} -\Expe{\rvax^{j}+\rvaz^{j}|\rvax^{k}})}
=
\Expe{\rvax(i)(\rvaz^{j} -\Expe{\rvaz^{j}|\rvax^{k}})}
$.
Since $\rvaz^{\ell}_{G},\rvax^{\ell}$ have the same joint second-order statistics as 
$\rvaz^{\ell},\rvax^{\ell}$, it follows that
$
\Expe{\rvax(i)(\rvay_{G}^{j} -\Expe{\rvay_{G}^{j}|\rvax^{k}})}
=0,\, \forall j\leq k<i\leq N$.
This, together with the fact that $\rvaz_{G}^{\ell}$ is jointly Gaussian with $\rvax^{\ell}$,
implies that also the reconstructed sequence $\set{\rvay_{G}(k)}\eq \set{\rvax(k)+\rvaz_{G}(k)}$
satisfies the causality constraint~\eqref{eq:Markov_Causal}.

We are now in the position to state the first main result of this section:
 \begin{lem}\label{lem:italwaysbounds}
For any zero-mean Gaussian random vector source of length $\ell$ having bounded differential entropy, and for every $D>0$,
\begin{align}
 R_{c}^{op(\ell)}(D) \leq R_{c}^{it(\ell)}(D)+\frac{1}{2}\log_{2}(2\pi \expo{}) \fspace \text{bits/sample}.
\end{align}
\finenunciado
\end{lem}
The proof of Lemma~\ref{lem:italwaysbounds} is presented in Section~\ref{sec:proof_italwaysbounds}.

The result stated in Lemma~\ref{lem:italwaysbounds} for Gaussian random vector sources is extended to Gaussian stationary processes in the following theorem (the second main result of this section):
\begin{thm}\label{thm:italwaysboundsproc}
 For a zero-mean Gaussian stationary source $\procx$, and $D>0$,
 \begin{align}
  R_{c}^{op}(D) \leq R_{c}^{it}(D) + \frac{1}{2}\log_{2}(2\pi\expo{}).
 \end{align}
\finenunciado
\end{thm}
The proof of Theorem~\ref{thm:italwaysboundsproc} can be found in Section~\ref{sec:proof_italwaysboundsproc}.

The fact that $R_{c}^{it}(D)+(1/2)\log_{2}(2\pi\expo{} )\geq R_{c}^{op}(D)$ for Gaussian sources allows one to find upper bounds to the OPTA of causal codes by explicitly finding or upper bounding $R_{c}^{it}(D)$.
This is accomplished in the following sections.

\section{$R_{c}^{it}(D)$ for 
First-Order Gauss-Markov Processes}\label{sec:Rcit(D)_1st_order}

In this section we will find $R_{c}^{it}(D)$ when the source is a first-order Gauss-Markov process.
More precisely, we will show that the information-theoretic causal RDF $R_{c}^{it}(D)$, which is associated with
an average distortion constraint, coincides with the expression for the SRDF on the RHS of~\eqref{eq:Rsrd} obtained
in~\cite{tatiko00} for a per-sample distortion constraint.
To do so, and to provide also a constructive method of realizing the SRDF as well as $R_{c}^{it}(D)$, we will start by stating an alternative derivation of the SRDF defined in~\cite{tatiko00}.
%

Before proceeding, it will be convenient to introduce some additional notation.
For any process $\procx$, we write $\rvex^{j}_{k}$, $j\leq k$, to denote the random column vector $[\rvax(j)\cdots \rvax(k)]^{T}$ and adopt the shorter notation $\rvax_{k}\eq \rvax(k)$.
For any two random vectors $\rvex^{j}_{k}$, $\rvey^{\ell}_{m}$, we define
$\bK_{\rvex^{j}_{k}}\eq \Expe{\rvex^{j}_{k}(\rvex^{j}_{k})^{T}}$,
$\bK_{\rvey^{\ell}_{m}\rvex^{j}_{k}}\eq \Expe{\rvey^{\ell}_{m}(\rvex^{j}_{k})^{T}}$.
%


It was already stated in Lemma~\ref{lem:ZisGaussian} that the reconstruction process $\rvay^{\ell}$ which realizes mutual information for any given MSE distortion constraint,  must be jointly Gaussian with the source. 
This holds in particular for a realization of the SRDF with distortion schedule $D_{1},\ldots,D_{\ell}$.
In the next theorem we will obtain an explicit expression for this RDF and prove that in its realization, the sample distortions $\Expe{(\rvay(k)-\rvax(k))^{2}}$ equal the \emph{effective distortions} $\set{d_{k}}_{k=1}^{\ell}$, defined as 
\begin{subequations}\label{eq:sigsqz_must_equal}
\begin{align}
 d_{1}
 &\eq
 \min\Set{\sigsq_{\rvax(1)}\,,\,D_{1}}
 \\
  d_{k}
&\eq 
 \min\Set{a_{k-1}^{2} d_{(k-1)} +\sigsq_{\xi(k-1)} \,,\, D_{k}  }
 ,\fspace\forall k=2,\ldots \ell.
\end{align}
\end{subequations}
Moreover, it will be shown that the 
unique second-order statistics of this realization are given by the following recursive algorithm:

\vspace{0mm}
\framebox{
\begin{minipage}{0.9\linewidth}
\emph{
\begin{center}
	\textbf{Procedure 1}
\end{center}
}
\vspace{-10mm}
$ $\\
Step 0: Set $\Expe{\rvay_{1}^{2}}=\Expe{\rvay_{1}\rvax_{1}}=\Expe{\rvax_{1}^{2}}-d_{1}$. 
\\
Step 1: Set the counter $k=2$.
\\
Step 2: Set $\Expe{\rvey^{1}_{k-1}\rvax_{k}} = 
 \bK_{\rvey^{1}_{k-1}\rvex^{1}_{k-1}}(\bK_{\rvex^{1}_{k-1}})^{-1} \Expe{\rvex^{1}_{k-1}\rvax_{k}}$
 \\
Step 3: Set $\Expe{\rvey^{1}_{k-1}\rvay_{k}} =\Expe{\rvey^{1}_{k-1}\rvax_{k}}$
\\
Step 4: Set $\Expe{\rvay_{k}^{2}}=\Expe{\rvay_{k}\rvax_{k}}=\Expe{\rvax_{k}^{2}}-d_{k}$
\\
Step 5: Enlarge $\bK_{\rvey^{1}_{k-1}}$ to $\bK_{\rvey^{1}_{k}}$ by appending the column $\Expe{\rvey^{1}_{k-1}\rvay_{k}}$ and the row $\Expe{\rvey^{1}_{k}\rvay_{k}}^{T}$, calculated in steps 3 and 4.
\\
Step 6: Set $\Expe{\rvay_{k}(\rvex^{1}_{k-1})^{T}}$ as  
 \begin{align}\label{eq:Past_indep_trans}
\Expe{\rvay_{k}(\rvex^{1}_{k-1})^{T}}
 &=
 \expe{\rvey^{1}_{k} \rvax_{k}}^{T}
 \left( 
 \begin{matrix}
  \bK_{\rvey^{1}_{k-1}} & \Expe{\rvey^{1}_{k-1} \rvax_{k}}\\
  \Expe{\rvey^{1}_{k-1} \rvax_{k}}^{T} & \sigsq_{\rvax_{k}}
 \end{matrix}
 \right)^{-1}
 \left[
 \begin{matrix}
 \bK_{\rvey^{1}_{k-1}\rvex^{1}_{k-1}}
 \\
 \Expe{\rvex^{1}_{k-1} \rvax_{k}}^{T} 
 \end{matrix}
  \right]  
 \end{align}
 \\
Step 7: 
Put together $\bK_{\rvey^{1}_{k-1}\rvex^{1}_{k-1}}$, $\Expe{\rvay_{k}\rvex^{1}_{k-1}}$,  $\Expe{\rvey^{1}_{k-1}\rvax_{k}}$ and $\Expe{\rvay_{k}\rvax_{k}}$ to obtain $\bK_{\rvey^{1}_{k}\rvex^{1}_{k}}$.
\\
Step 8: Increment $k$ by~$1$ and go to Step~2.
\end{minipage}
}
$ $\\

\begin{figure}[htbp]
 \centering
 \input{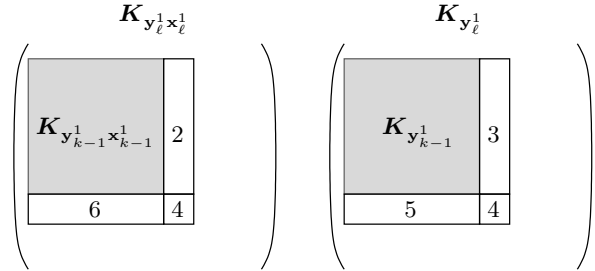}
  \caption{Illustration of the recursive Procedure~1 at its $k$-th iteration.
 Starting from known covariance matrices $\bK_{\rvey^{1}_{k-1}\rvex^{1}_{k-1}}$, $\bK_{\rvey^{1}_{k-1}}$, their next partial rows and columns are found.
 The numbers indicate the step in the algorithm which reveals the corresponding part of the matrix.}
\label{fig:Matrices}
\end{figure}

Figure~\ref{fig:Matrices} illustrates the operation of the above recursive procedure.
After $k-1$ iterations, the covariance sub-matrices $\bK_{\rvey^{1}_{k-1}\rvex^{1}_{k-1}}$, $\bK_{\rvey^{1}_{k-1}}$
have been found.
At the $k$-th iteration, step~$i$ is responsible of revealing the partial rows and columns indicated by number~$i$ in the figure.

The above results are formally stated in the following theorem, which also gives an exact expression for the SRDF of first-order Gauss-Markov sources.
%
\begin{thm}\label{thm:longone}
 Let $\set{\rvax(k)}_{k=1}^{\ell}$ be a first-order Gauss-Markov source of the form
 \begin{align}\label{eq:thesource}
  \rvax(k+1) = a_{k}\rvax(k) + \xi(k), \fspace k=1,\ldots,\ell-1,
 \end{align}
where $\rvax(1)$  and the innovations $\set{\xi(k)}_{k=1}^{\ell-1}$ are independent zero-mean Gaussian random variables with variances $\sigsq_{\rvax(1)}$ and $\set{\sigsq_{\xi(k)}}_{k=1}^{\ell-1}$, respectively.
Then, the sequential rate distortion function (SRDF) for $\set{\rvax(k)}_{k=1}^{\ell}$ under distortion schedule 
 $\set{D_{k}}_{k=1}^{\ell}$ is given by
 \begin{align}\label{eq:therightexpression}
  R^{it}_{SRD}(D_{1},\ldots,D_{\ell})
  =
  \frac{1}{2\ell}
\ln\left(\frac{\sigsq_{\rvax(1)}}{d_{1}} \right)
+
\frac{1}{2\ell}
\sumfromto{k=2}{\ell}
\ln\left( 
\frac{a_{k-1}^{2} d_{k-1} +\sigsq_{\xi(k-1)}  }{d_{k}}
\right),
 \end{align}
where the \emph{effective distortions} $\set{d_{k}}_{k=1}^{\ell}$ are
defined in~\eqref{eq:sigsqz_must_equal}.
The unique second-order statistics of a realization of $R_{c}^{it}(D)$ for this source are obtained by the recursive algorithm described in Procedure~1.
\finenunciado
\end{thm}

The proof of this theorem can be found in Section~\ref{sec:proof_longone}.

\begin{rem}
The expression for the SRDF with per-sample distortion constraints in~\eqref{eq:therightexpression} differs from the one found in~\cite[p.~186]{tatiko00} for the source~\eqref{eq:thesource} with $a_{k}=a$, $\forall k=1,\ldots,\ell$, which in our notation reads
\begin{align}\label{eq:Tatikonda's_SRDF_expr}
 R_{\ell}^{SRD}(D_{1},\ldots,D_{\ell})
 =
 \frac{1}{\ell}\Sumfromto{t=1}{\ell}
 \max\Set{0\,,\, \frac{1}{2}\log\left( \frac{a^{2}D_{t-1} + \sigsq_{\xi(t-1)} }{D_{t}}\right) },
\end{align}
wherein $D_{0}=0$ and $\sigsq_{\xi(0)}=\sigsq_{\rvax(1)}$.
The difference lies in that 
the logarithms in~\eqref{eq:therightexpression} contain the \textit{effective } distortions $\set{d_{k}}_{k=1}^{\ell}$, whereas~\eqref{eq:Tatikonda's_SRDF_expr}  uses the distortion constraints $\set{D_{k}}_{k=1}^{\ell}$ themselves.
It is likely that the author of~\cite{tatiko00}, on page 186, intended these distortion constraints to be the effective distortions, i.e.,  that $\Expe{(\rvay(k)-\rvax(k))^{2}}=D_{k}$, for every $k=1,\ldots,\ell$.
However, on~\cite[Definition~5.3.5 on p.~147]{tatiko00}, the SRDF under a distortion schedule is defined as the infimum of a mutual information rate subject to the constraints $\Expe{(\rvay(k)-\rvax(k))^{2}}\leq D_{k}$.
Under the latter interpretation, nothing precludes one from choosing an arbitrarily large value for, say, $D_{1}$, yielding an arbitrarily large value for the second term in the summation on the RHS of~\eqref{eq:Tatikonda's_SRDF_expr}, which is, of course, inadequate.
\finenunciado
\end{rem}

We are now in a position to find the expression for $R_{c}^{it}(D)$ for first-order Gauss-Markov sources.
This is done in the following theorem, whose proof is contained in Section~\ref{sec:proof_shortone}.
\begin{thm}\label{thm:shortone}
For a stationary Gaussian process
\begin{align}
 \rvax(k+1)
 =
 a
 \rvax(k)
 +
 \xi(k),\fspace k=1,\ldots 
\end{align}
where $\set{\xi(k)}$ is an i.i.d. sequence of zero-mean Gaussian random variables with variance $\sigsq_{\xi}$,
$\rvax(1)\sim N(0,\sigsq_{\rvax})$ with~$\sigsq_{\rvax}\eq \sigsq_{\xi}/(1-a^{2})$,
the information-theoretic causal RDF is given by
\begin{align}\label{eq:R_{c}^{it}(D)_explicit}
 R_{c}^{it}(D)
 =
 \frac{1}{2}\ln\left(a^{2} +\frac{\sigsq_{\xi}}{D} \right).
\end{align}
\finenunciado
\end{thm}


The technique applied to prove theorems~\ref{thm:longone} and~\ref{thm:shortone}  does not seem to be extendable to
Gauss-Markov processes of order greater than~1.
In the sequel,
we will find upper bounds to $R_{c}^{it}(D)$ for arbitrary (any order) stationary Gaussian
sources.

\section{Closed-Form Upper Bounds}\label{sec:Analytic_Bounds}
In order to upper bound the difference between $R_{c}^{it}(D)$ and $R(D)$ for arbitrary stationary Gaussian sources, we will start this
section by
defining an upper bounding function for $R_{c}^{it}(D)$, denoted by
$\overline{R_{c}^{it}}(D)$.
We will then derive three closed-form upper bounding functions to the rate-loss
$\overline{R_c^{it}}(D)-R(D)$, applicable to any Gaussian stationary process.
Two of these bounds are strictly smaller than $0.5$ bit/sample for all distortions
$0<D\leq \sigsq_{\rvax}$.

We begin with the following definition:
\begin{defn}[Causal Stationary RDF]\label{def:Stat_CausalRDF}
	For a stationary source~$\procx$, the information-theoretic Causal Stationary Rate-Distortion
Function $\overline{R_{c}^{it}}(D)$ is defined as
	\begin{align*}
		\overline{R_{c}^{it}}(D) \eq \inf\bar{I}(\procx;\procy),
	\end{align*}
where the infimum is over all processes $\procy$ such that: 
\begin{itemize}
	\item[i)] $d(\procx,\procy)\leq D$,
	\item[ii)] the reconstruction error $\procz \eq \procy - \procx$ is jointly stationary with the
source, and
	\item[iii)] Markov chain~\eqref{eq:Markov_Causal} holds.
\end{itemize}
\finenunciado
\end{defn}

Next we derive three closed-form upper bounding functions to $\overline{R_{c}^{it}}(D)-R(D)$ that
are applicable to arbitrary zero-mean stationary Gaussian sources with finite differential entropy rate.
This result
is stated in the following
theorem, proved in Section~\ref{sec:proof_Bouds_B}:
%
\begin{thm}\label{thm:Bouds_B}
\emph{
Let $\procx$ be a zero-mean Gaussian stationary source with PSD $S_{\rvax}\ejw$ with bounded differential entropy rate and variance
$\sigsq_{\rvax}$.
Let $R(D)$ denote Shannon's RDF for $\procx$ (given by~\eqref{eq:R(D)waterfill}),
and let $R^{\perp}(D)$ denote the quadratic Gaussian RDF for source-uncorrelated distortions for the
source $\procx$ defined in~\eqref{eq:Rperpdef}.
Let $R_{c}^{it}(D)$ denote the information-theoretic causal RDF (see
Definition~\ref{def:CausalRDF}).
Then, for all $D\in(0,\sigsq_{\rvax})$,
\begin{align}\label{eq:ineqs_B}
R_{c}^{it}(D) - R(D) \leq 
\overline{R_c^{it}}(D)-R(D) 
\leq
B_{1}(D)
\leq
B_{2}(D)
<
B_{3}(D)
\leq 0.5 \textrm{ bits/sample},
\end{align}
where 
\begin{align}
B_{1}(D) & \eq R^{\perp}(\tfrac{\sigsq_{\rvax} D}{\sigsq_{\rvax} -D}) - R(D)\label{eq:B2}
\\
B_{2}(D) & \eq \frac{1}{4\pi}
\intfromto{-\pi}{\pi}
\log_{2} \left(1 + [1 -\tfrac{D}{\sigsq_{\rvax}}]\frac{S_{\rvax}\ejw   }{D}\right)d\w -
R(D)\label{eq:B3}
\\
B_{3}(D) &\eq
\min\Set{ 
\frac{1}{2}\log_{2}
   \left((1+\tfrac{\vareps}{D})
   \left[  1 + (\varsigma_{\rvax}^{\vareps} -\tfrac{1}{\sigsq_{\rvax}})D \right]\right)
   \,,\,
   0.5
   \,,\,
   \frac{1}{2}\log_{2}\left( \frac{\sigsq_{\rvax}}{D}\right)
},\label{eq:B4}
\end{align}
 where
 \begin{align}\label{eq:theintegral}
  \varsigma_{\rvax}^{\vareps}\eq
  \intpipi{\frac{1}{\max\Set{\vareps,S_{\rvax}\ejw}}},
 \end{align}
with 
$\vareps$ being any non-negative scalar with which~\eqref{eq:theintegral} exists and such that $\vareps\leq D$. 
}
\finenunciado
\end{thm}

Notice that $B_{3}(D)$ is independent of $R(D)$, being therefore numerically simpler to evaluate
than the other bounding functions introduced in Theorem~\ref{thm:Bouds_B}.
However, as $D$ is decreased away from $\sigsq_{\rvax}$ and approaches $\sigsq_{\rvax}/2$ ,
$B_{3}(D)$ becomes very loose.
In fact, it can be seen from~\eqref{eq:B4a} that for $D>\sigsq_{\rvax}/2$, the gap between
$R_{c}^{it}(D)$ and $R(D)$ is actually upper bounded by $B_{3}(D)-R(D)$, which is of course tighter
than $B_{3}(D)$, but requires one to evaluate $R(D)$.

It is easy to see that time-sharing between two causal realizations with distortions $D_{1}$, $D_{2}$ and rates $R_{c}^{it}(D_{1})$, $R_{c}^{it}(D_{2})$ yields an output process which satisfies causality with a rate-distortion pair corresponding to the linear combination of $R_{c}^{it}(D_{1})$, $R_{c}^{it}(D_{2})$.
Thus, in some cases one could get a bound tighter than $B_{3}$ by considering the boundary of the convex hull of the region above $R(D)+B_{3}(D)$ and then subtracting $R(D)$.
However, such bound would be much more involved to compute, since it requires to evaluate not only $R(D)$, but also the already mentioned convex hull.

It is also worth noting that the first term within the $\min$ operator on the RHS of~\eqref{eq:B4} 
becomes smaller when $\varsigma^{\vareps} - 1/\sigsq_{\rvax}$ is reduced.
This difference, which from Jensen's inequality is always non-negative, could be taken as a measure of the ``non-flatness'' of the PSD of $\procx$ (specially when $\vareps=0$).
Indeed, as $\procx$ approaches a white process, $B_{3}$ tends to zero.

It can be seen from~\eqref{eq:ineqs_B} that $\overline{R_{c}^{it}}(D)$ provides the tightest upper
bound for the information-theoretic RDF among all bounds presented so far.
Although it does not seem to be feasible to obtain a closed-form expression for
$\overline{R_{c}^{it}}(D)$, we show in the next section how to get arbitrarily close to
it.

\section{Obtaining $\overline{R_{c}^{it}}(D)$}\label{sec:convexity}
In this section we present an iterative procedure that allows one to 
calculate $\overline{R_{c}^{it}}(D)$ with arbitrary accuracy, for any $D>0$.
In addition, we will see that this procedure yields a characterization of the filters in a dithered feedback
quantizer~\cite{dersil08} that achieve
an operational rate which is upper bounded by $\overline{R_{c}^{it}}(D)+0.254$~[bits/sample].

\subsection{An Equivalent Problem}
\begin{figure}[htbp]
\centering
\input{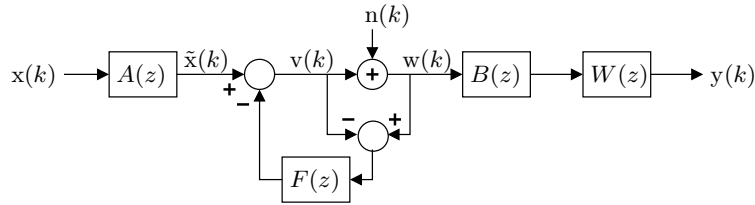}
\caption{AWGN channel within a ``perfect reconstruction'' system followed by causal de-noising
filter $W(z)$.}
\label{fig:causalFQ}
\end{figure}

To derive the results mentioned above, we will work on a scheme consisting of an AWGN channel and a set of causal
filters, as depicted in Fig.~\ref{fig:causalFQ}.
In this scheme, the source $\procx$ is Gaussian and stationary, with PSD $S_{\rvax}\ejw$, and is
assumed to have finite differential entropy rate.
In Fig.~\ref{fig:causalFQ}, the noise $\procn$ is a zero-mean Gaussian process with i.i.d. samples,
independent of $\procx$.
Thus, between $\rvav(k)$ and $\rvaw(k)$ lies the AWGN channel $\rvaw(k) = \rvav(k)+\rvan(k)$.
The filter $F(z)$ is stable and strictly causal, i.e., it has at least a one sample delay.
The filters $A(z)$ and $B(z)$ are causal and stable.
The idea, to be developed in the remainder of this section, is to first show that with the filters
that minimize the variance of the reconstruction error for a fixed ratio
$\sigsq_{\rvaw}/\sigsq_{\rvan}$, the system of Fig.~\ref{fig:causalFQ} attains a mutual information rate between source and reconstruction equal to
$\overline{R_{c}^{it}}(D)$, with a reconstruction MSE equal to $D$.
We will then show that finding such filters is a convex optimization problem, which naturally
suggests an iterative procedure to solve it.

In order to analyze the system in Fig.~\ref{fig:causalFQ}, and for notational convenience,
we define
\begin{align*}
	\Omx\ejw \eq \hsqrt{S_{\rvax}\ejw}, \fspace \forallwinpipi.
\end{align*}
We also restrict the filters $A(z)$ and $B(z)$ to satisfy the ``perfect reconstruction'' condition
\begin{align}\label{eq:PR_causal}
	A\ejw B\ejw \equiv 1.
\end{align}
Thus,
\begin{align}\label{eq:y(k)}
	\rvay(k) =     W(z) \rvax(k) + W(z)B(z)[1-F(z)] \rvan(k),
\end{align}
see Fig.~\ref{fig:causalFQ}.
Therefore, $W(z)$ is the signal transfer function of the system.

\begin{figure}[h!]
 \centering
 \input{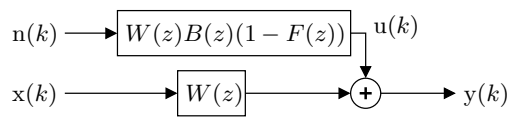}
\caption{Equivalent block diagram depicting the output as the sum of $W(z)x(k)$ and $u(k)$, where $\procn$ is an i.i.d. zero-mean Gaussian process independent of $\procx$.}
\label{fig:roles}
\end{figure}

The perfect reconstruction condition~\eqref{eq:PR_causal} induces a division of roles in the system,
which will later translate into a convenient parametrization of the optimization problem associated
with it.
On the one hand, because of~\eqref{eq:PR_causal}, the net effect of the AWGN channel and the filters
$A(z)$, $B(z)$ and $F(z)$ is to introduce (coloured) Gaussian stationary additive noise, namely $\procu$, independent
of the source.
The PSD of this noise, $S_{\rvau}\ejw$, is given by
\begin{align}\label{eq:Su_def}
	S_{\rvau}\ejw \eq \abs{W \ejw}^{2}\abs{B\ejw}^{2}\abs{1-F\ejw}^{2} \sigsq_{\rvan}.
\end{align}
The diagram in Figure~\ref{fig:roles} shows how the signal transfer function $W(z)$ and the noise transfer function $W(z)B(z)(1-F(Z))$ act upon $\procx$ and $\procn$ to yield the output process.

On the other hand, by looking at Fig.~\ref{fig:causalFQ} one can see that $W(z)$ plays also the role of a de-noising filter, which can be utilized to reduce
additive noise at the expense of introducing linear distortion.
More precisely, $W(z)$ acts upon the Gaussian stationary source $\procx$ corrupted by additive
Gaussian stationary noise with PSD $\abs{B\ejw}^{2}\abs{1-F\ejw}^{2}\sigsq_{n}$.
From~\eqref{eq:y(k)} and Fig.~\ref{fig:causalFQ}, the MSE is given by
\begin{align}\label{eq:Dc_expr}
	D_{c} 
	\eq
	\sigsq_{\rvau} + \norm{(W-1)\Omx}^{2}
	=
	\frac{\norm{\Omx A}^{2} \norm{WBf }^{2} }{K-\norm{f}^{2}} + \norm{(W-1)\Omx}^{2},
\end{align}
where $\sigsq_{\rvau}\eq \intpipi{S_{\rvau}\ejw}$ and
\begin{align*}
	K			&\eq \frac{\sigsq_{\rvav}}{\sigsq_{\rvan}} +1 =
\frac{\sigsq_{\rvaw}}{\sigsq_{\rvan}},\\
	f(\w) 		&\eq \abs{1-F\ejw},\fspace \forallwinpipi.
\end{align*}
On the RHS of~\eqref{eq:Dc_expr}, the first term is the variance of the additive, source
independent, Gaussian noise.
The second term corresponds to the error due to linear distortion, that is, from the deviation of
$W\ejw$ from a unit gain.

Since we will be interested in minimizing $D_{c}$, for any given $F(z)$ and $W(z)$, the filters
$A(z)$ and $B(z)$ in Fig.~\ref{fig:causalFQ} are chosen so as to
minimize~$\sigsq_{\rvau}$ in~\eqref{eq:Dc_expr}, while still satisfying~\eqref{eq:PR_causal}.
From the viewpoint of the subsystem comprised of the filters $A(z)$, $B(z)$ and $F(z)$ and the AWGN
channel, $W(z)$ acts as an error frequency weighting filter, see~\eqref{eq:Su_def}.
Thus, for any $F(z)$ and $W(z)$, the filters $A(z)$ and $B(z)$ that minimize $\sigsq_{\rvau}$ are
those characterized in~\cite[Prop.~1]{dersil08}, by setting $P(z)$ in~\cite[eq.~(20b)]{dersil08}
equal to $W(z)$.
With the minimizer filters in~\cite{dersil08}, the variance of the source-independent error term is
given by
\begin{align}\label{eq:sigsq_u}
	\sigsq_{\rvau} = \frac{\ip{\Omx \abss{W},  f }^{2}}{K-\norm{f}^{2}}.
\end{align}
On the other hand, the filter $F(z)$ needs to be strictly causal and stable.
As a consequence, it holds that 
\begin{align*}
	\Intfromto{-\pi}{\pi}\log f(\w) d\w  \geq 0,
\end{align*}
which follows from Jensen's formula~\cite{carkro83} (see also the Bode Integral Theorem in,
e.g.,~\cite{serbra97}).

Thus, from~\eqref{eq:Dc_expr} and~\eqref{eq:sigsq_u}, if one wishes to minimize the reconstruction
MSE by choosing appropriate \emph{causal} filters in the system in Fig.~\ref{fig:causalFQ} for a
given value of $K$, one needs to solve the following optimization problem:
\begin{opprob}\label{opprob:fW}
	For any  given $\Omx\ejw$,  and for any given $K>1$, find the frequency response $W\ejw$ and the
frequency response magnitude $f(\w)$ that
	\begin{subequations}
	\begin{align}
		\textrm{Minimize: 		}\; &  D_{c}\eq \frac{\ip{\Omx \abss{W},  f }^{2}}{K-\norm{f}^{2}} +
\norm{(W-1)\Omx}^{2} \label{eq:Dc_def}\\
		\textrm{Subject to: 	}\; &  W\in\Hset,\nonumber\\
											&  \intfromto{-\pi}{\pi} \ln f(\w) d\w\geq0,\nonumber
	\end{align}
	\end{subequations}
	where $\Hset$ denotes the space of all frequency responses that can be realized with causal
filters.
	\finenunciado
\end{opprob}

Now we can establish the equivalence between solving Optimization Problem~\ref{opprob:fW} and
finding $\overline{R_{c}^{it}}(D)$.
\begin{lem}\label{lem:filters_realize_Rc}
	For any $K>1$ and $\Omx\ejw$, if the filters $A^{\star}(z)$, $B^{\star}(z)$, and $F^{\star}(z)$ 
solve Optimization Problem~\ref{opprob:fW} and yield distortion $D_{c}^{\star}$, then
	\begin{align*}
		\frac{1}{2}\ln(K) = \overline{R_{c}^{it}}(D_{c}^{\star}).
	\end{align*}
\finenunciado
\end{lem}

From the above lemma, whose 
proof can be found in Section~\ref{sec:proof_filters_realize_Rc},
one can find $\overline{R_{c}^{it}}(D)$ either by solving the minimization in
Definition~\ref{def:Stat_CausalRDF} or by solving Optimization Problem~\ref{opprob:fW}.
In the following, we will pursue the latter approach.
As we shall see, our formulation of Optimization Problem~\ref{opprob:fW} provides a convenient
parametrization of its decision variables.
In fact, it makes it possible to establish the convexity of 
the cost functional defined in~\eqref{eq:Dc_def} with respect to the set of all causal frequency
responses involved.
That result can be obtained directly from the following key lemma, proved in Section~\ref{sec:proof_Jsc()_is_convex}:
\begin{lem}\label{lem:Jsc(fg)_is_convex}
	Define the sets of functions
	\begin{align*}
		\Fset_{K}	&\eq \Set{f:\pipi\to\Rl^{+}_{0}, \norm{f}^{2} < K},\\
		\Gset	&\eq \Set{G:\pipi\to \mathbb{C} },
	\end{align*}
where $K$ is some positive constant.
	Then, for any $G\in\Gset$ and $K>1$, the cost functional $\Jsc: \Fset_{K}\times\Gset \to
\Rl^{+}_{0}$, defined as
	\begin{align}\label{eq:V2}
		\Jsc(f,g)\eq \frac{\ip{f,\abs{g}}^{2}}{K-\norm{f}^{2}} + \norm{g-G}^{2},
	\end{align}
	is strictly convex in $f$ and $g$.
\finenunciado
\end{lem}

We can now prove the convexity of Optimization Problem~\ref{opprob:fW}.
\begin{lem}\label{lem:Jsc()_is_convex}
	For all $\Omx$ and for all $K>1$, Optimization Problem~\ref{opprob:fW} is convex .
	\finenunciado
\end{lem}
\begin{proof}
	With the change of variables $G\eq \Omx$ and $g\eq \Omx W$ in~\eqref{eq:V2}, we obtain
$D_{c}=\Jsc(f,g)$, see~\eqref{eq:Dc_expr}.
	With this, Optimization Problem~\ref{opprob:fW} amounts to finding the functions $f$ and $g$
that
	\begin{subequations}\label{eq:opprob_fg}
	\begin{align}
		\textrm{Minimize: }\;		&\Jsc(f,g)\\
		\textrm{Subject to: }\;	& g\in\Wset,f\in\Bset.
	\end{align}
	\end{subequations}
	where 
	\begin{align}
	\Wset 	&\eq\set{g = \Omx W:  W\in\Hset}\label{eq:Wset_def}\\
	\Bset	&\eq\Set{f\in\Fset_{K}: \intfromto{-\pi}{\pi}\ln f(\w) d\w =0 }.\nonumber
	\end{align}
	Clearly, the space of frequency responses associated with causal transfer functions, $\Hset$, is
a convex set.
	This implies that $\Wset$ is a convex set.
	In addition, $\Bset$ is also a convex set, and from Lemma~\ref{lem:Jsc(fg)_is_convex},
$\Jsc(f,g)$ is a convex functional.
	Therefore, the optimization problem stated in~\eqref{eq:opprob_fg}, and thus Optimization
Problem~\ref{opprob:fW},
	are convex. This completes the proof.
\end{proof}

\subsection{Finding $\overline{R_{c}^{it}}(D)$ Numerically}
Lemma~\ref{lem:Jsc()_is_convex} and the parametrization in Optimization Problem~\ref{opprob:fW} 
allow one to define an iterative algorithm that, as will be shown later, yields the
information-theoretic causal RDF.
Such algorithm is embodied in iterative Procedure~2:

$ $\\
\framebox{
\begin{minipage}{0.9\linewidth}
\emph{
\begin{center}
	\textbf{Iterative Procedure 2}
\end{center}
For any target information theoretical rate $R$, 
\begin{enumerate}
	\item[] Step 1: Set $K=2^{2R}$.
	\item[] Step 2: Set $W\ejw\equiv 1$.
	\item[] Step 3: Find the frequency response magnitude $f\in\Bset$ that minimizes $D_{c}$ for
given $W$.
	\item[] Step 4: Find the causal frequency response $W\in\Hset$ that minimizes $D_{c}$ for given
$f$.
	\item[] Step 5: Return to step 3.\\
\end{enumerate}
}
\end{minipage}
}\label{anchor}

$ $

Notice that after  solving Step 3 in the first iteration of Procedure~2, the MSE is
comprised of only additive noise independent of the source.%
\footnote{Indeed, after solving Step 3 for the first time, the resulting rate is the quadratic
Gaussian rate distortion function for source uncorrelated distortions $R^{\perp}(D)$ introduced
in~\cite{derost08} (see also~\eqref{eq:R^{perp}}) .}
Step 4 then reduces the MSE by attenuating source-independent noise at the expense of introducing
linear distortion.
Each step reduces the MSE until a local (or global) minimum of the MSE is obtained.
Based upon the convexity of Optimization Problem~\ref{opprob:fW}, the following theorem, which is
the main technical result in this section, guarantees convergence to the global minimum of the MSE,
say $D$, for a given end-to-end mutual information.
Since all the filters in Optimization Problem~\ref{opprob:fW} are causal, 
the mutual information achieved at this global minimum is equal to $\overline{R_{c}^{it}}(D)$.
\begin{thm}[Convergence of iterative Procedure 2]\label{thm:convergence}
Iterative Procedure~2 converges  monotonically to the unique $f$ and $W$ that realize
$\overline{R_{c}^{it}}(D)$.
More precisely, letting $\Delta^{(n)}$ denote the MSE obtained after the $n$-th iteration of Iterative
Procedure~2 aimed at a target rate $R$, we have that%
\begin{align*}
 n_{2}>n_{1} \iff \Delta^{(n2)}< \Delta^{(n1)}
\end{align*}
and
\begin{align*}
 \lim_{n\to\infty}\overline{R_{c}^{it}}(\Delta^{(n)})=R.
\end{align*}
\finenunciado
\end{thm}

\begin{proof}
	The result follows directly from the fact that Optimization Problem~\ref{opprob:fW} is strictly convex in
$f$ and $W$, which was shown in Lemma~\ref{lem:Jsc(fg)_is_convex}, and from
Lemma~\ref{lem:filters_realize_Rc}.
\end{proof}
The above theorem states that the stationary information-theoretic causal RDF can be obtained by
using Iterative Procedure~2.
In practice, this means that an approximation arbitrarily close to $\overline{R_{c}^{it}}(D)$ for a
given $D$ can be obtained if sufficient iterations of the procedure are carried out.

The feasibility of running Iterative Procedure~2 depends on being able to solve each of the
minimization sub-problems involved in steps~3 and~4.
We next show how these sub-problems can be solved.

\subsection*{Solving Step 3}
If $W\ejw$ is given, the minimization problem in Step~3 of Iterative Procedure~2 is equivalent to
solving a feedback quantizer design problem with the constraint $A(z)B(z)=1,\, \forall z\in
\mathbb{C}$ and with error weighting filter $W\ejw$.
Therefore, the solution to Step~3 is given in closed form by~\cite[eqs.~(20),~(29) and (31b)]{dersil08}, where
$P(z)$ in~\cite[eq.~(20b)]{dersil08} is replaced by $W(z)$.
The latter equations of~\cite{dersil08} characterize the frequency response magnitudes of the optimal $A(z)$, $B(z)$
and $1-F(z)$ given $W(z)$.
The existence of rational transfer functions $A(z)$, $B(z)$ and $F(z)$ arbitrarily close (in an $\Ldos$ sense) to
such frequency response magnitudes is also shown in~\cite{dersil08}. 
\subsection*{Solving Step 4}\label{step4}
Finding the causal frequency response $W\ejw\in\Hset$ that minimizes $D_{c}$ for a given $f$ is
equivalent to
solving
\begin{align}\label{eq:minimize_over_g}
	\min_{g: g\in\Wset}\Jsc(f,g)
\end{align}
for a given $f$, where $\Wset$ is as defined in~\eqref{eq:Wset_def}.
Since $\Wset$ and $\Jsc(\cdot,\cdot)$ are convex,~\eqref{eq:minimize_over_g} is a convex
optimization problem.
As such, its global solution can always be found iteratively.
In particular, if $W(z)$ is constrained to be an $M$-th order FIR filter with impulse response
$\bc\in\Rl^{M+1}$, such that $W\ejw= \Fsp\Set{\bc}$, where $\Fsp\set{\cdot}$ denotes the
discrete-time Fourier transform, then
\begin{align*}
	\Gsc(\bc) \eq \Jsc(f, \Fsp\set{\bc}) 
\end{align*}
is a convex functional.
The latter follows directly from the convexity of $\Jsc(\cdot,\cdot)$ and 
the linearity of~$\Fsp\set{\cdot}$.
As a consequence, one can solve the minimization problem in Step 4, to any degree of accuracy, by
minimizing $\Gsc(\bc)$ over the values of the impulse response of $W\ejw$, using standard convex
optimization methods (see, e.g,~\cite{boyvan04}).
This approach also has the benefit of being amenable to numerical computation.

It is interesting to note that if the order of the de-noising filter $W(z)$ were not a priori
restricted, then, after Iterative Procedure~2 has converged to $\overline{R_{c}^{it}}(D)$, the
obtained $W(z)$ is the causal Wiener filter (i.e., the MMSE causal estimator)
for the noisy signal that comes out of the perfect reconstruction system that precedes $W(z)$.
Notice also that one can get the system in Fig.~\ref{fig:filters} to yield a realization of
Shannon's $R(D)$ using Iterative Procedure~1 by simply allowing $W(z)$ to be non-causal.
This would yield a system equivalent to the one that was obtained analytically in~\cite{zamkoc08}.
An important observation is that one could not obtain a realization of $R_{c}^{it}(D)$ from
such a system in one step by simply replacing $W(z)$ (a non-causal Wiener filter) by the MMSE causal
estimator (that is, a causal Wiener filter).
To see this, it suffices to notice that, in doing so, the frequency response magnitude of $W(z)$
would change.
As a consequence, the previously matched filters $A(z)$, $B(z)$ and $F(z)$ would no longer be
optimal for $W(z)$.
One would then have to change $A(z)$, and then $W(z)$ again, and so on, thus having to carry out
infinitely many recursive optimization steps.
However, a causally truncated version of the non causal Wiener filter $W(z)$ that realizes  Shannon's
RDF could be used as an alternative starting guess in Step~2 of the iterative procedure.

\subsection{Achieving $\overline{R_{c}^{it}}(D)+0.254$ bits/sample Causally}\label{subsec:Rop_causal}
\begin{figure}[htbp]
 \centering
 \input{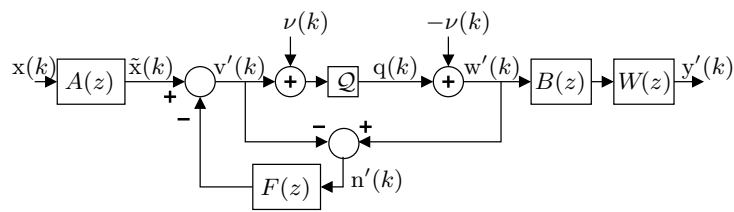}
  \caption{Uniform scalar quantizer $\Qsp$ and dither signals $\nu(k)$, $-\nu(k)$, forming an SDUSQ, replacing the AWGN channel of the system from Fig.~\ref{fig:causalFQ}.%
\label{fig:SDUSQ_with_filters}
 }
\end{figure}

If the AWGN channel in the system of Fig.~\ref{fig:causalFQ} is replaced by a \emph{subtractively
dithered uniform scalar quantizer} (SDUSQ), as shown in Fig.~\ref{fig:SDUSQ_with_filters},
then instead of the noise $\procn$ we will have an
i.i.d. process independent of $\procx$, whose samples are uniformly distributed over the
quantization interval~\cite{zamfed96b}.
The dither signal, denoted by $\set{\nu(k)}$, is an i.i.d. sequence of uniformly distributed random
variables, independent of the source.
Let $\proc{q}$ be the quantized output of the SDUSQ.
Denote the resulting input and the output to the quantizer, before adding and after subtracting the
dither, respectively, as  $\set{\rvav'(k)}$ and  $\set{w'(k)}$, and let $\set{\rvan'(k)}\eq
\set{\rvaw'(k)-\rvav'(k)}$ be the quantization noise introduced by the SDUSQ.
Notice that the elements of $\set{\rvan'(k)}$ are independent, both mutually and from the source
$\procx$.
However, unlike $\procv$ and $\procw$, the processes $\set{\rvav'(k)}$ and $\set{\rvaw'(k)}$ are not
Gaussian, since they contain samples of the uniformly distributed process $\set{\rvan'(k)}$.
We then have the following:
\begin{thm}\label{thm:causal_bound_with_ECDQ}
If the scheme shown in Fig.~\ref{fig:SDUSQ_with_filters} uses the filters yielded by Iterative Procedure~2, and if long sequences of the quantized output of this system are entropy coded conditioned to the dither values in a memoryless fashion, then an operational rate $r_{c}^{op}$ satisfying
\begin{align}
 r_{c}^{op}\leq \overline{R_{c}^{it}}(D) + \tfrac{1}{2}\log_{2}(2\pi\expo{})
\end{align}
is achieved causally while attaining a reconstruction MSE equal to $D$.
\finenunciado
\end{thm}
\begin{proof}
If memoryless entropy coding is applied to long sequences of symbols conditioning the probabilities to dither values, then then operational rate equals the conditional entropy $ H(\rvaq(k)|\nu(k)) $.
For this entropy, the following holds in the system shown in Fig.~\ref{fig:SDUSQ_with_filters}:
\begin{equation}\label{eq:rate_loss}
\begin{split}
 H(\rvaq(k)|\nu(k)) 
&
\overset{(a)}{=} I(\rvav'(k);\rvaw'(k))
= 
I(\rvav'(k);\rvav'(k)+\rvan'(k)) 
\overset{\hphantom{(b)}}{=} h(\rvav'(k)+\rvan'(k)) - h(\rvan'(k))\\
&\overset{(b)}{=} h(\rvav(k)+\rvan(k)) - h(\rvan(k)) + D(\rvan'(k)\Vert \rvan(k)) -
D(\rvav'(k)+\rvan'(k)\Vert \rvav(k) + \rvan(k))\\
&\overset{\hphantom{(b)}} {<} I(\rvav(k);\rvav(k)+\rvan(k)) + D(\rvan'(k)\Vert \rvan(k))
= 
I(\rvav(k);\rvaw(k)) + \tfrac{1}{2}\log_{2}(\tfrac{2\pi\expo{}}{12})
\\&= 
\tfrac{1}{2}\log_{2}K + \tfrac{1}{2}\log_{2}(\tfrac{2\pi\expo{}}{12})
\end{split}
\end{equation} 
where $H(\rvaq(k)|\nu(k))$ denotes the entropy of $\rvaq(k)$ conditioned to the $k$-th value of the
dither signal.
In the above,~$(a)$ follows from~\cite[Theorem~1]{zamfed92}. 
In turn,~$(b)$ stems from the well known result $\Dsp(\rvax'\Vert \rvax) = h(\rvax) - h(\rvax')$, where
$\Dsp(\cdot\Vert \cdot)$ denotes the Kullback-Leibler distance,  see,
e.g.,~\cite[p.~254]{covtho06}.
The inequality in the last line of~\eqref{eq:rate_loss} is strict since the distribution of
$\rvav'(k)$  is not Gaussian.

The result follows directly by combining~\eqref{eq:rate_loss}  
with  Lemma~\ref{lem:filters_realize_Rc} and Theorem~\ref{thm:convergence}. 
\end{proof}

In view of Theorem~\ref{thm:causal_bound_with_ECDQ}, and since any ED pair using an SDUSQ and LTI filters yields a reconstruction error jointly stationary with the source, it follows that the operational rate-distortion performance of the feedback quantizer thus obtained is within 
$0.5\log_{2}\left(2\pi \text{e}/12\right)\simeq 0.254$
bits/sample from the best performance achievable by any ED pair within this class.

\begin{rem}\label{rem:highrate}
When the rate goes to infinity, so does $K$.
In that limiting case, the transfer function $W(z)$ tends to unity, and 
it follows from~\cite{dersil08} that the optimal filters asymptotically satisfy
$\abs{A\ejw}=S_{\rvax}\ejw^{-1}$,
$\abs{B\ejw}=S_{\rvax}\ejw$,
$\abs{1-F\ejw}=\exp\left({\intpipi{\ln(S_{x}\ejw)}}\right)/S_{x}\ejw$.
Moreover, when $K\to\infty$, the system of Fig.~\ref{fig:SDUSQ_with_filters} achieves $R_{c}^{op}(D)$ which, in this asymptotic regime, coincides with $\overline{R_{c}^{it}}(D)+0.5\log_{2}(2\pi\text{e})$, with $\overline{R_{c}^{it}}(D)$ tending to $R(D)$.
\finenunciado
\end{rem}

\subsection{Achieving $\overline{R_{c}^{it}}(D)+1.254$ bits/sample With Zero Delay}\label{subsec:Rop_zd}
If the requirement of zero-delay, which is stronger than that of causality, was to be satisfied,
then it would not be possible to apply entropy coding to long sequences of quantized samples.
This would entail an excess bit-rate not greater than $1$ bit per sample, see,
e.g.,~\cite[Section~5.4]{covtho06}.
Consequently, we have the following result:

\begin{thm}\label{thm:zd_bound_with_ECDQ}
The OPTA of zero-delay codes, say $R_{ZD}^{op}(D)$, can be upper bounded by the operational rate of the scheme of Fig.~\ref{fig:SDUSQ_with_filters} when each quantized output value is entropy-coded independently, conditioned to the current dither value. 
Thus
\begin{align}\label{eq:Rzd}
R_{ZD}^{op}(D) \leq \overline{R_{c}^{it}}(D) +\frac{1}{2}\ln\left(\frac{2\pi e}{12}\right)
+1
\simeq
\overline{R_{c}^{it}}(D) + 0.254 +1\fspace \textrm{bits/sample}.
\end{align}
 \finenunciado
\end{thm}

The $0.254$ bits per sample in~\eqref{eq:Rzd}, commonly referred to as the
``space-filling loss'' of scalar quantization, can be reduced by using vector
quantization~\cite{gispie68,zamfed92}.
Vector quantization could be applied while preserving causality (and without introducing delay) if
the samples of the source were $N$-dimensional vectors.
This would also allow for the use of entropy coding over $N$-dimensional vectors of quantized
samples, which reduces the extra $1$ bit/sample at the end of~\eqref{eq:Rzd} to $1/N$ bits/sample,
see~\cite[Theorem~5.4.2]{covtho06}.

\subsection{The Additive Rate Loss of Causality Arises from Two Factors}
It is worth noting that Lemma~\ref{lem:filters_realize_Rc} and the above analysis reveals an
interesting fact: the rate loss due to causality for Gaussian sources with memory, that is, the
difference between the OPTA of causal codes and $R(D)$, is upper bounded by the sum of two terms.
The first term is $0.254$ bits/sample, and results from the space filling loss associated with
scalar quantization, as was also pointed out in~\cite{linzam06} for the high resolution situation.
This term is associated only with the \emph{encoder}.
For a scalar Gaussian stationary source, such excess rate can only be avoided by jointly quantizing
blocks of consecutive source samples (vector quantization), i.e., by allowing for non-causal
encoding (or by encoding several parallel sources).
The second term can be attributed to the reduced de-noising capabilities of causal filters, compared
to those of non-causal (or smoothing) filters.
The contribution of the causal filtering aspect to the total rate-loss is indeed
$\overline{R_{c}^{it}}(D)-R(D)$.
This latter gap can also be associated with the performance loss of causal \emph{decoding}.

As a final remark, we note that the architecture of Fig.~\ref{fig:causalFQ}, which allowed us to
pose the search of $R_{c}^{it}
(D)$ as a convex optimization problem, is by no means the only scheme capable of achieving the upper
bounds~\eqref{eq:rate_loss} and~\eqref{eq:Rzd}.
For instance, it can be shown that the same performance can be attained removing either $A(z)$ or
$F(z)$ in the system of Fig.~\ref{fig:causalFQ}, provided an entropy coder with infinite memory is
used.
Indeed, the theoretical optimality (among causal codes) of the differential pulse code
modulation (DPCM) architecture, with predictive feedback and causal MMSE estimation at the decoding
end, has been shown in a different setting~\cite{mawish07}.

\section{Example}\label{sec:Example}
To illustrate the upper bounds presented in the previous sections,
we here evaluate $B_{1}(D)$, $B_{2}(D)$, and $B_{3}(D)$, and calculate an approximation of
$\overline{R_{c}^{it}}(D)$ via Iterative Procedure~2, for two Gaussian zero-mean  AR-1 and AR-2 sources.
These sources were generated by the recursion
%
%
\begin{align}\label{eq:recursion}
 \rvax(k) = a_{1} \rvax(k-1) + a_{2} \rvax(k-2) + \rvaz(k),\fspace\forall k\in\Z,
\end{align}
where the elements of the process $\procz$ are i.i.d. zero-mean unit-variance Gaussian random
variables.

Iterative Procedure~2 was carried out by restricting $W(z)$ to be an 8-tap FIR filter.
For each of the target rates considered, the procedure was stopped after four complete iterations.

\begin{figure}[htb]
 \centering
\input{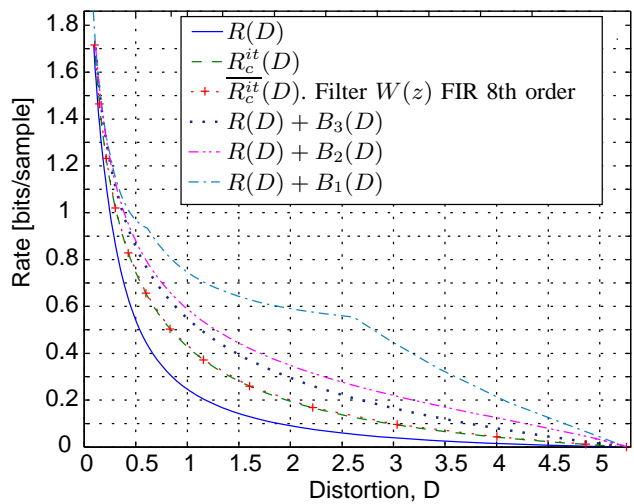}
\caption{R(D) (in bits/sample) and several upper bounding functions for $R_{c}^{it}(D)$ for zero-mean unit variance
white Gaussian noise filtered through $z/(z-0.9)$.
The resulting source variance is $5.26$.}
 \label{fig:bounds_09}
\end{figure}

\begin{figure}[htb]
 \centering
\input{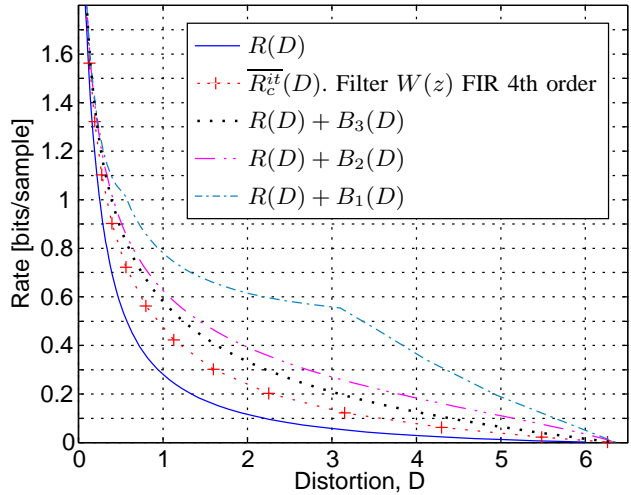}
 \caption{R(D) (in bits/sample) and several upper bounding functions for $R_{c}^{it}(D)$ for zero-mean unit variance
white Gaussian noise filtered through $z^{2}/[(z-0.9)(z-0.1)]$.
The resulting source variance is $6.37$.}
 \label{fig:bounds_0901}
\end{figure}

The first-order source (Source~1) was chosen by setting the values of the coefficients in~\eqref{eq:recursion} to be
$a_{1}=0.9$, $a_{2}=0$.
This amounts to zero-mean, unit variance white Gaussian noise filtered through the colouring transfer function
$z/(z-0.9)$.
The second-order source (Source~2) consisted of zero-mean, unit variance white Gaussian noise filtered through the
colouring transfer function
$z^{2}/[(z-0.9)(z-0.1)]$.
The resulting upper bounds for  Source~1 and Source~2 are shown in Figs.~\ref{fig:bounds_09} and~\ref{fig:bounds_0901}, 
respectively.
As predicted by~\eqref{eq:Boudsineqs} and~\eqref{eq:B4}, all the upper bounds for $R_{c}^{it}(D)$
derived in Section~\ref{sec:Analytic_Bounds} converge to $R(D)$ in the limit of both large and small
distortions (that is, when $D\to {\sigsq_{\rvax}}^{-}$ and $D\to 0^{+}$, respectively).

For both sources, the gap between $\overline{R_{c}^{it}}(D)$ and $R(D)$ is significantly smaller than $0.5$ bits/sample, 
for all rates at which $\overline{R_{c}^{it}}(D)$ was evaluated.
Indeed, this gap is smaller than $0.22$ bit/sample for both sources.

\begin{figure}[htb]
 \centering
 \includegraphics[width=0.55\columnwidth]{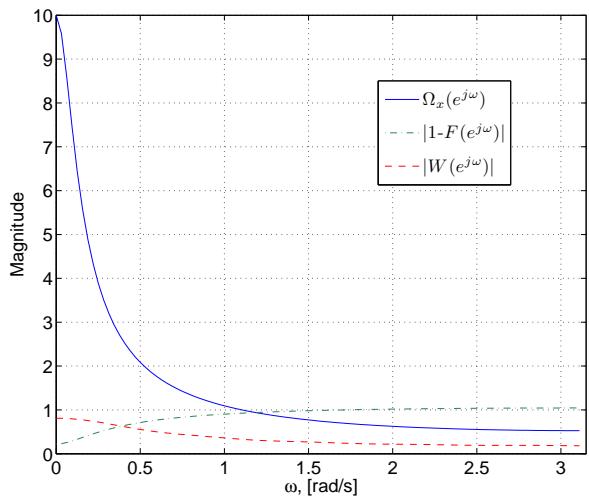}
 \caption{$\Omx\ejw$, $\abs{1-F\ejw}$ and $\abs{W\ejw}$ of an approximate realization of $\overline{R_{c}^{it}}(D)$ for a Gaussian stationary source with 
PSD $\abs{1/(1-0.9\expo{-j\w})}^{2}$ when the rate is $0.2601$ [bit/sample], using the system shown in
Fig.~\ref{fig:causalFQ}.
These frequency responses were obtained after four iterations of Iterative Procedure~1, with filter $W(z)$ being FIR
with 8 taps.}
 \label{fig:filters}
\end{figure}

For the first-order source, the magnitude of the coefficients of the FIR filter $W(z)$ obtained decays rapidly with 
coefficient index.
For example, when running five cycles of Iterative Procedure~2, using a 10th order FIR filter for~$W(z)$, 
for Source~1 at $R=0.2601$ bits/sample, the obtained $W(z)$  was
\begin{multline*}
W(z)=
0.3027+
0.1899z^{-1}+
0.1192z^{-2}+
0.0748z^{-3}+
0.0470z^{-4}+
0.0296z^{-5}+
0.0188z^{-6}\\+
0.0123z^{-7}+
0.0086z^{-8}+
0.0070z^{-9} 
\end{multline*}
Such fast decay of the impulse response of $W(z)$ suggests that, at least for AR-1 sources, there is little to be gained by letting 
$W(z)$ be an FIR filter of larger order.
(It is worth noting that, in the iterative procedure, the initial guess for $W(z)$ is a unit scalar gain.)
The frequency response magnitude of $W(z)$ is plotted in Fig.~\ref{fig:filters}, together with $\Omx\ejw$ and the
resulting frequency response magnitude
$\abs{1-F\ejw}$ after four iterations on Source~1 for a target rate
of $\overline{R_{c}^{it}}(D)=0.2601$ bits/sample.

Notice that for Source~1, after four iterations of Iterative Procedure~1, the obtained values for
$\overline{R_{c}^{it}}(D)$ are almost identical to $R_{c}^{it}(D)$, evaluated according to~\eqref{eq:R_{c}^{it}(D)_explicit}.
This
suggests 
that Iterative Procedure~2 has fast convergence.
For example, when applying four iterations of Iterative Procedure~2 to Source~1 with a target rate of $0.2601$ bits/sample, 
the distortions obtained after each iteration were $1.6565$, $1.6026$, $1.6023$ and $1.6023$, respectively.
For the same source with a target rate of $0.0441$ bits/sample, the distortion took the values 
$4.0152$, $3.9783$,    $3.9783$, and   $3.9782$ as the iterations proceeded.
A similar behaviour is observed for other target rates, and for other choices of $a_{1}$ in~\eqref{eq:recursion} as well.
Thus, at least for AR-1 sources, one gets close to the global optimum $\overline{R_{c}^{it}}(D)$ after just three iterations.

\section{Conclusions}\label{sec:concl}
In this paper we have obtained expressions and upper bounds to the causal and zero-delay rate distortion function for Gaussian stationary sources and MSE as the distortion measure.
We first
showed that for Gaussian sources with bounded differential entropy rate, the causal OPTA does not exceed the information-theoretic RDF by more than approximately~$0.254$ bits/sample.
After that, we derived an explicit expression for the information-theoretic RDF under per-sample MSE distortion constraints using a constructive method.
This result was then utilized for obtaining
a closed-form formula for the causal information-theoretic RDF $R_{c}^{it}(D)$ of first-order Gauss-Markov
sources under an average MSE distortion constraint.

We then derived three closed-form upper bounding functions to the 
difference between $R_{c}^{it}(D)$ and Shannon's RDF. 
Two of these bounding functions are
tighter than the previously best known bound of $0.5$  bits/sample, at all
rates.
We also provided a tighter fourth upper bound to $R_c^it(D)$, named
$\overline{R_c^{it}}(D)$,  that is constructive.
More precisely, we provide a practical scheme that  attains this bound,
based on a noise-shaped predictive coder consisting of  an AWGN channel
surrounded by pre-, post-, and feedback  filters.
For a given source spectral density  and
desired distortion, the design of the filters is convex in  their
frequency responses. 
We proposed an iterative algorithm,  which is
guaranteed to converge to the optimal set of unique  filters. 
Moreover, the mutual information obtained across the AWGN  channel, converges
monotonically to $\overline{R_c^{it}}(D)$.
Thus, one avoids having to solve the more complicated
minimization of the mutual information over
all possible conditional distributions satisfying the distortion
constraint.
To achieve the upper bounds on the operational coding rates, one may simply replace
the AWGN channel by a subtractively-dithered scalar quantizer and using memoryless entropy coding conditioned to the dither values.

\section{Proof of Lemma~\ref{lem:italwaysbounds}}\label{sec:proof_italwaysbounds}
We will first show that $R_{c}^{it(\ell)}(D)$ can be realized by a vector AWGN channel between two square matrices.
It was already established in Lemma~\ref{lem:ZisGaussian} that an output $\rvey$ corresponds to a realization of $R_{c}^{it(\ell)}$ only if it is jointly Gaussian with the source $\rvex$.
From this Gaussianity condition, the MMSE estimator of $\rvey$ from $\rvex$, 
say $\hat{\rvey}$, is given by
\begin{align}\label{eq:hatrvey}
 \hat{\rvey} = \bK_{\rvey \rvex}\bK_{\rvex}^{-1}\rvex,
\end{align}
where the inverse of $\bK_{\rvex}$ exists from the fact that $\rvex$ has bounded differential entropy.
It is clear from~\eqref{eq:hatrvey} and the joint Gaussianity between $\rvex$ and $\rvey$ that the causality condition is satisfied if and only if the matrix 
\begin{align}\label{eq:matrix_LT}
\bK_{\rvey \rvex}\bK_{\rvex}^{-1} \;\text{ is lower triangular.}
\end{align}
On the other hand, the distortion constraint~\eqref{eq:dconstrell} can be expressed as
\begin{align}\label{eq:dintermsofKs}
\tfrac{1}{\ell} \tr{\Expe{(\rvey-\rvex)(\rvey-\rvex)^{T} }}
=
\tfrac{1}{\ell}
\tr{\bK_{\rvey} -\bK_{\rvey\rvex} - (\bK_{\rvey\rvex})^{T}+\bK_{\rvex}}
\leq D
\end{align}
From the definition of $R_{c}^{it(\ell)}$, for every $\vareps>0$, there exists an output vector $\rvey$ jointly Gaussian with $\rvex$ such that $\bK_{\rvey}$ and $\bK_{\rvey\rvex}$ satisfy~\eqref{eq:matrix_LT},~\eqref{eq:dintermsofKs} and
\begin{align}\label{eq:IleqRpluseps}
 \frac{1}{\ell}I(\rvex;\rvey)\leq R_{c}^{it(\ell)}+\vareps.
\end{align}
We will now describe a simple scheme which is capable of reproducing the joint statistics between $\rvex$ and any given $\rvey$ jointly Gaussian with $\rvex$ satisfying~\eqref{eq:matrix_LT},~\eqref{eq:dintermsofKs} and~\eqref{eq:IleqRpluseps}. 

Suppose $\rvex$ is first multiplied by a matrix $\bA\in\Rl^{\ell\times \ell}$ yielding the random vector $\rvev\eq \bA\rvex$.
Then a vector with Gaussian i.i.d. entries with unit variance, independent from $\rvex$, say $\rven \in\Rl^{\ell}$, is added to $\rvev$, to yield the random vector $\rvew\eq \rvev +\rven$.
Finally, this result is multiplied by a matrix $\bB\in\Rl^{\ell\times \ell}$ to yield the output
\begin{align}
\rvey= \bB \bA \rvex  + \bB \rven .
\end{align}
On the other hand, the joint second-order statistics between $\rvey$ and $\rvex$ are fully characterized by the matrices
\begin{align}
\bK_{\rvey\rvex}&
=
\Expe{\rvey\rvex^{T}} 
= 
\bB\bA\bK_{\rvex}
\\
\bK_{\rvey}
&
= 
\Expe{\rvey \rvey^{T}}
=
\bB \bA \bK_{\rvex}(\bB\bA)^{T}
+
\bB \bB^{T}.
\end{align}
It can be seen from these equations that all that is needed for the system described above to reproduce any given pair of covariance matrices 
$\bK_{\rvey}$, $\bK_{\rvey\rvex}$  
is that the matrices $\bA $ and $\bB $ satisfy
\begin{align}
\bB  \bA 
 &=  
\bK_{\rvey \rvex }
\bK_{\rvex }^{-1}\label{eq:AB}
\\
\bB \bB^{T}
&=
\bM\eq 
\bK_{\rvey} - \bK_{\rvey\rvex}\bK_{\rvex}^{-1}\bK_{\rvex\rvey}\label{eq:BBT}
\end{align}
Thus, $\bB$ can be chosen, for example, as the lower-triangular matrix in a Cholesky factorization of $\bM$.
With this, a tentative solution for $\bA$ could be obtained as 
$\bA=\bB^{\dagger}\bK_{\rvey\rvex}\bK_{\rvex}^{-1}$, which would satisfy~\eqref{eq:AB} if and only if
$ 
\bB\bB^{\dagger}\bK_{\rvey\rvex}\bK_{\rvex}^{-1} = \bK_{\rvey\rvex}\bK_{\rvex}^{-1}
$.
The latter holds if and only if 
$\ospn\set{\bK_{\rvey\rvex}}\subseteq\ospn\set{\bB}$ (recall that $\bK_{\rvex}$ is non-singular since $\rvex$ has bounded differential entropy).
We will now show that this condition actually holds by using a contradiction argument.
Suppose $\ospn\set{\bK_{\rvey\rvex}} \nsubseteq\ospn\set{\bB}$.
Since $\ospn\set{\bB}=\ospn\set{\bM}$, the former supposition is equivalent to $\ospn\set{\bK_{\rvey\rvex}} \nsubseteq\ospn\set{\bM}$. 
If this were the case, then there would exist $\bs\in\Rl^{\ell}$ such that 
$\bs^{T}\bK_{\rvey\rvex}\neq \bzero$ and $\bs^{T}\bM=\bzero$.
The latter, combined with~\eqref{eq:BBT}, would imply $\bs^{T}\bK_{\rvey}\neq 0$.
One could then construct the scalar random variable $\rvar\eq \bs^{T}\rvey$, which would have non-zero variance.
The MSE of predicting $\rvar$ from $\rvex$ is given by 
$$
\bK_{\rvar} 
-
\bK_{\rvar\rvex}\bK_{\rvex}^{-1}\bK_{\rvar\rvex}^{T}
=
\bs^{T}
(\bK_{\rvey}
-
\bK_{\rvey\rvex}\bK_{\rvex}^{-1}\bK_{\rvex\rvey}
)\bs
=
\bs^{T}\bM\bs
=
0.
$$
From this, and in view of the fact that $\rvar$ is Gaussian with non-zero variance, we conclude that $I(\rvex;\rvar)$ would be unbounded.
However, by construction, the Markov chain $\rvar \leftrightarrow \rvey \leftrightarrow\rvex$ holds, and therefore by the Data Processing Inequality we would have that $I(\rvex;\rvey)\geq I(\rvex;\rvar)$, implying that $I(\rvex;\rvey)$ is  unbounded too.
This contradicts the assumption that $\rvey$ is a realization of $R_{c}^{it(\ell)}(D)$, leading to the conclusion that $\ospn\set{\bK_{\rvey\rvex}}\subseteq \ospn\set{\bB}$.
Therefore, the choice 
\begin{align}\label{eq:AN}
\bA=\bB^{\dagger}\bK_{\rvey\rvex}\bK_{\rvex}^{-1}
\end{align}
is guaranteed to satisfy~\eqref{eq:AB}, and thus for every $\vareps>0$, there exist matrices $\bB$ and $\bA$ which yield an output vector satisfying~\eqref{eq:matrix_LT},~\eqref{eq:dintermsofKs} and~\eqref{eq:IleqRpluseps}.

On the other hand, we have that
\begin{align}
 I(\rvex ;\rvey )
 =
I(\rvev ;\rvey )
=
I(\rvev ;\rvew )
\end{align}
The first equality follows from the data-processing inequality and the fact that $\rvev$ is obtained deterministically from $\rvex$.
The second equality stems from~\eqref{eq:AN}, which implies that $\ospn\set{\bA}\cap\Nsp\set{\bB}= \emptyset$.
The latter means that $\bB$ is invertible along all the directions in which $\rvev$ has energy, which together with the fact that $\rven $ is i.i.d. and independent of $\rvev $ implies
$h(\rvev |\rvey ) 
=
h(\rvev |\rvew )$.
Therefore, if $\bA $ and $\bB $ yield an output $\rvey$ such that $(1/\ell)I(\rvex;\rvey)\leq R_{c}^{it(\ell)}(D)+\vareps$, then 
$
\frac{1}{\ell}I(\rvev;\rvew)
\leq
R_{c}^{it(\ell)}(D) +\vareps
$.

Finally, if we keep the $\bA$ and $\bB$ satisfying the above conditions and replace the noise $\rven $ by the vector of noise samples $\rvem $ with unit variance introduced by $\ell$ independently operating subtractively-dithered uniform scalar quantizers (SDUQS)~\cite{zamfed92}, with their outputs being jointly entropy-coded conditioned to the dither, then  the operational data rate 
$r(\rvex,\rvey)\eq \Expe{\rvaL_{\ell}(\rvex)}$ would be upper bounded by~\cite{zamfed92}
\begin{align*}
 r(\rvex,\rvey) \leq  \bar{I}(\rvev ;\rveu ) + \frac{1}{\ell}
 \leq 
 + \bar{I}(\rvev ;\rvew ) 
 +
 \frac{1}{2}\log_{2}(2\pi\expo{})
 + \frac{1}{\ell}
\end{align*}
where $\rveu \eq \rvev +\rvem $ is the output of the ECDQ channel.
Since the distortion yielded by the SDUQs is the same as that obtained with the original Gaussian channel, we conclude that 
\begin{align*}
 R_{c}^{op(\ell)}(D) 
 \leq
 R_{c}^{it(\ell)}(D)+ \frac{1}{2}\log_{2}(2\pi\expo{})
 +
 \frac{1}{\ell}
 +\vareps
 \fspace \text{bits/sample}.
\end{align*}
Given that the above holds for any $\vareps>0$ and since $R_{c}^{op(\ell)}(D)$ is defined as an infimum, we conclude that  
$ R_{c}^{op(\ell)}(D)\leq
 R_{c}^{it(\ell)}(D)+ \frac{1}{2}\log_{2}(2\pi\expo{})
 +
 \frac{1}{\ell}$,  
which completes the proof.
\findemo


\section{Proof of Theorem~\ref{thm:italwaysboundsproc} }\label{sec:proof_italwaysboundsproc}
We will start by showing that
\begin{align}
 R_{c}^{op}(D) = \limsup_{\ell\to\infty}R_{c}^{op(\ell)}(D).
\end{align}
First, following exactly the same proof as in Lemma~\ref{lem:inflingeqliminf} in the Appendix, it is straightforward to show that 
\begin{align}\label{eq:LlimRop}
 R_{c}^{op}(D) \geq \limsup_{\ell\to\infty}R_{c}^{op(\ell)}(D).
\end{align}
Now, consider the following family of encoding/decoding schemes.
For some positive integer $\ell$, the entire source sequence is encoded in blocks of $\ell$ contiguous samples.
Encoding and decoding of each block is independent of the encoding and decoding of any other block.
As in the scheme described in the second part of the proof of Lemma~\ref{lem:italwaysbounds}, each block is encoded and decoded utilizing $\ell$ parallel and independent SDUSQs, with their outputs jointly entropy coded conditioned to the dither values, and using with the optimal pre- and post-processing matrices.
For such an ED pair, and from~\eqref{eq:r_def}, the operational rate after $k$ samples have been reconstructed is
\begin{align}\label{eq:r<Rcop+algo}
r(\rvax^{k},\rvay^{k}) 
= 
\frac{\ell}{k}\left\lceil \frac{k}{\ell} \right\rceil R_{c}^{op(\ell)}(D)
<
R_{c}^{op(\ell)}(D)
+
\frac{\ell}{k}
R_{c}^{op(\ell)}(D),
\end{align}
where $\lceil\cdot \rceil$ denotes rounding to the nearest larger integer
(since the $k$-th sample is reconstructed only after $\lceil k/\ell \rceil$ blocks of length $\ell$ are decoded).
On the other hand, since the variance of each reconstruction error sample cannot be larger than the variance of the source, we have that the average distortion associated with the first $k$ samples is upper bounded as
\begin{align}
 d(\rvax^{k},\rvay^{k})
 \leq
 \frac{\ell \lfloor k/\ell \rfloor}{k} D  
 +
 \frac{k -\ell \lfloor k/\ell \rfloor }{k} \sigsq_{\rvax}
 <
 D
 +
 \frac{\ell}{k}\sigsq_{\rvax},
\end{align}
where $\lfloor\cdot \rfloor$ denotes rounding to the nearest smaller integer.
Therefore, for any finite $\ell$, the average distortion of this scheme equals $D$ when $k\to\infty$ (i.e., when we consider the entire source process).
Also, from~\eqref{eq:r<Rcop+algo} and~\eqref{eq:r_def},
letting $k\to\infty$ we conclude that
\begin{align}
 R_{c}^{op}(D)
\leq 
 R_{c}^{op(\ell)}(D).
\end{align}
If $\limsup_{\ell}R_{c}^{op(\ell)}(D)$ exists, then, for every $\vareps>0$, there exists a finite $\ell_{0}(\vareps)\in\Nl$ such that
\begin{align}
 R_{c}^{op(\ell)}(D)\leq \limsup_{\ell}R_{c}^{op(\ell)}(D) +\vareps,\fspace \forall\ell \geq \ell_{0}(\vareps)
\end{align}
Therefore, every $\vareps>0$, there exists a finite $\ell_{0}(\vareps)\in\Nl$ such that
\begin{align}\label{eq:Rcopleqlim+espilon}
  R_{c}^{op}(D)\leq \limsup_{\ell\to\infty} R_{c}^{op(\ell)}(D) +\vareps,\fspace \forall \ell\geq \ell_{0}(\vareps)
\end{align}
Since $R_{c}^{op}(D)$ is defined as an infimum among all causal codes (which, in particular, means $\ell$ can be chosen larger than $\ell_{0}(\vareps)$ for any $\vareps>0$), it readily follows  from~\eqref{eq:LlimRop},~\eqref{eq:Rcopleqlim+espilon}, Lemma~\ref{lem:italwaysbounds} and Lemma~\ref{lem:limusp}, that
\begin{align*}
  R_{c}^{op}(D)
  = \limsup_{\ell\to\infty} R_{c}^{op(\ell)}(D)
  \leq \limsup_{\ell\to\infty} R_{c}^{it(\ell)}(D) + \frac{1}{2}\log_{2}(2\pi\expo{} )
  \leq
  R_{c}^{it}(D) + \frac{1}{2}\log_{2}(2\pi\expo{} ),
\end{align*}
completing the proof.
\findemo


\section{Proof of Theorem~\ref{thm:longone}}\label{sec:proof_longone}
From Lemma~\ref{lem:ZisGaussian},
 for any given reconstruction-error covariance matrix, the mutual information is minimized if and only if the output is jointly Gaussian with the source.
In addition, for any given mutual information between $\rvax^{\ell}$ and a jointly Gaussian output $\rvay^{\ell}$, the variance 
of every reconstruction error sample $\rvaz(k)\eq \rvay(k)-\rvax(k)$ is minimized if and only if 
$\rvaz(k)$ is the estimation error resulting from estimating $\rvax(k)$ from $\rvay^{k}$, that is, if and only if
\begin{align}\label{eq:z_perp_y}
 0=\Expe{\rvaz_{k}\rvey^{1}_{k}}=\Expe{(\rvay_{k}-\rvax_{k})\rvey^{1}_{k}},\fspace\forall k=1,\ldots, \ell,
\end{align}
which for Gaussian vectors implies $\rvaz(k)$ and $\rvay^{k}$ are independent, and therefore
\begin{align}\label{eq:z_indep_y}
 h(\rvaz(k)|\rvay^{k}) =h(\rvaz(k)),\fspace \forall k=1,\ldots, \ell.
\end{align}
Thus, hereafter we restrict the analysis to output processes jointly Gaussian with and causally related to $\rvax^{\ell}$ which also satisfy~\eqref{eq:z_perp_y}.
For any such output process, say, $\rvay^{\ell}$, the following holds:
\begin{align}
  I(\rvax^{\ell};\rvay^{\ell})
  &=
\frac{1}{\ell}
\sumfromto{k=1}{\ell}I(\rvax^{\ell};\rvay(k)|\rvay^{k-1})\nonumber
\\&
=
\frac{1}{\ell}
\sumfromto{k=1}{\ell}I(\rvax^k;\rvay(k)|\rvay^{k-1})\label{eq:estaotra}
\\&
\geq 
\frac{1}{\ell}
\sumfromto{k=1}{\ell}I(\rvax(k);\rvay(k)|\rvay^{k-1})\label{eq:thefirst}
\\&
=
\frac{
h(\rvax(1))
- 
h(\rvax(1)|\rvay(1))
}{\ell}
+
\frac{1}{\ell}
\sumfromto{k=2}{\ell}
\left[
h(\rvax(k)|\rvay^{k-1})
- 
h(\rvax(k)|\rvay^{k})\label{eq:lamedia}
\right]
\\&
=
\frac{
h(\rvax(1))
- 
h(\rvaz(1))
}{\ell}
+
\frac{1}{\ell}
\Sumfromto{k=2}{\ell}
\left[h(a_{k-1}\rvax(k-1)+\xi(k-1)|\rvay^{k-1})
- 
h(\rvaz(k))\label{eq:first}
\right]
\\&
=
\frac{1}{2\ell}
\ln\left(\frac{\sigsq_{\rvax(1)}}{\sigsq_{\rvaz(1)}} \right)
+
\frac{1}{\ell}
\sumfromto{k=2}{\ell}
\left[h(-a_{k-1}\rvaz(k-1)+\xi(k-1))
- 
h(\rvaz(k))
\right]\label{eq:refirst}
\\&
=
\frac{1}{2\ell}
\ln\left(\frac{\sigsq_{\rvax(1)}}{\sigsq_{\rvaz(1)}} \right)
+
\frac{1}{2\ell}
\sumfromto{k=2}{\ell}
\ln\left( 
\frac{a_{k-1}^{2} \sigsq_{\rvaz(k-1)} +\sigsq_{\xi(k-1)}  }{\sigsq_{\rvaz(k)}}
\right)\label{eq:Jane'saddiction}
\end{align}
In the above,~\eqref{eq:estaotra} follows because $\rvay^{\ell}$ depends causally upon $\rvax^{\ell}$.
In turn, inequality~\eqref{eq:thefirst} is due to the fact that 
$I(\rvax^{k};\rvay(k)|\rvay^{k-1})
= 
h(\rvay(k)|\rvay^{k-1})- h(\rvay(k)|\rvay^{k-1},\rvax^{k})
\geq
h(\rvay(k)|\rvay^{k-1})- h(\rvay(k)|\rvay^{k-1},\rvax(k))
$,
and thus equality holds in~\eqref{eq:thefirst} if and only if the following Markov chain is satisfied:
\begin{align}\label{eq:MC_key}
 \rvay(k)
 \leftrightarrow
 \set{\rvax(k),\rvay^{k-1} }
  \leftrightarrow
  \rvax^{k-1},\fspace \forall k=1,\ldots,\ell.
\end{align}
Finally,~\eqref{eq:first} and~\eqref{eq:refirst} follow because $\rvay^{\ell}$ satisfies~\eqref{eq:z_indep_y} for all $k=1,\ldots,\ell$.

Thus, the mutual information $I(\rvax^{\ell};\rvay^{\ell})$ of every output $\rvay^{\ell}$ that is a candidate to constitute a realization of $R_{\ell}^{SRD}(D_{1},\ldots,D_{\ell})$ is lower bounded by the RHS of~\eqref{eq:Jane'saddiction}, which in turn depends only on the error variances $\set{\sigsq_{\rvax(k)}}_{k=1}^{\ell}$ associated with $\rvay^{\ell}$.
We shall now see that this lower bound is minimized by a unique set of error variances, and then show that the resulting bound is achievable while having these error variances.

Revisiting~\eqref{eq:lamedia}~\eqref{eq:first} and~\eqref{eq:refirst},  we have that
$(1/2) \ln(
[a_{k-1}^{2} \sigsq_{\rvaz(k-1)} +\sigsq_{\xi(k-1)}  ]/\sigsq_{\rvaz(k)}
)
=
h(\rvax(k)|\rvay^{k-1})
-
h(\rvax(k)|\rvay^{k})
\geq 0
$
and
$
(1/2)\ln(\sigsq_{\rvax(1)}/\sigsq_{\rvaz(1)}) =h(\rvax(1)) -h(\rvax(1)|\rvay(1))\geq 0
$.
Therefore, in a realization of $R_{c}^{SRD}(D_{1},\ldots,D_{\ell})$, it holds that
\begin{subequations}\label{eq:subeq_eff_sigmas}
\begin{align}
\sigsq_{\rvaz(1)} &\leq \sigsq_{\rvax(1)}\\
 \sigsq_{\rvaz(k)}&\leq a_{k-1}^{2} \sigsq_{\rvaz(k-1)} +\sigsq_{\xi(k-1)}
 =
 \sigsq_{\rvax(k)} - a_{k-1}^{2}\sigsq_{\rvay(k-1)}
 ,\fspace\forall k=2,\ldots \ell.
\end{align}
\end{subequations}
With this, and since the right-hand side of~\eqref{eq:Jane'saddiction} decreases when any error variance~$\sigsq_{\rvaz(k)}$ increases, the minimum value of the right-hand side of~\eqref{eq:Jane'saddiction} subject to the constraints
\begin{align}\label{eq:sigsqz=d}
\sigsq_{\rvaz(k)}\leq D_{k},\; k=1,\ldots,\ell
\end{align}
is attained when these variances satisfy $\sigsq_{\rvaz(k)}=d_{k}$, for $k=1,\ldots,\ell$ (see~\eqref{eq:sigsqz_must_equal}).
Therefore, for all outputs $\rvay^{\ell}$ causally related to and jointly Gaussian with $\rvax^{\ell}$ satisfying the distortion constraints, it holds that
\begin{align}\label{eq:IgeqLB}
I(\rvax^{\ell};\rvay^{\ell})
\geq 
\frac{1}{2\ell}
\ln\left(\frac{\sigsq_{\rvax(1)}}{d_{1}} \right)
+
\frac{1}{2\ell}
\sumfromto{k=2}{\ell}
\ln\left( 
\frac{a_{k-1}^{2} d_{k-1} +\sigsq_{\xi(k-1)}  }{d_{k}}
\right),
\end{align}
with equality if and only if $\rvay^{\ell}$ satisfies~\eqref{eq:z_perp_y},~\eqref{eq:MC_key} and~\eqref{eq:sigsqz=d}.

Now we will show that for any distortion schedule $\set{D_{k}}_{k=1}^{\ell}$, the output $\rvay^{\ell}$ yielded by the recursive algorithm of Procedure~1 is such that $I(\rvax^{\ell};\rvay^{\ell})$ equals the lower bound~\eqref{eq:IgeqLB}, thus being a realization of $R_{\ell}^{SRD}(D_{1},\ldots,D_{\ell})$.

We will first demonstrate that $\procy$ satisfies the causality Markov chain
\begin{align}\label{eq:MC_causality}
\rvay_{k} 
  &
  \leftrightarrow
  \rvex^{1}_{k}
  \leftrightarrow
  \rvex^{k+1}_{\infty} \fspace \forall k\in\Nl
\end{align}
and the conditions%
~\eqref{eq:z_perp_y} (MMSE), 
and~\eqref{eq:MC_key} (Source's Past Independence)
which are necessary and sufficient to attain equality in~\eqref{eq:IgeqLB}.

\paragraph*{Causality condition~\eqref{eq:MC_causality}}
Let $\bA\eq \bK_{\rvey^{1}_{k}\rvex^{1}_{k}} (\bK_{\rvex^{1}_{k}})^{-1}$.
Suppose $\rvey^{1}_{k-1}$ satisfies causality. 
Then, since $\bK_{\rvey^{1}_{k}\rvex^{1}_{k}} = \bA\bK_{\rvex^{1}_{k}}$, it follows 
from~\eqref{eq:matrix_LT}
that
the top-left square submatrix $\bA^{k-1_{\!\lrcorner}}\in\Rl^{(k-1)\times (k-1)}$ of $\bA$ is lower triangular, being given by
\begin{align}
 \bA^{k-1_{\!\lrcorner}} = 
 \bK_{\rvey^{1}_{k-1}\rvex^{1}_{k-1}}(\bK_{\rvex^{1}_{k-1}})^{-1}.
\end{align}
Then Step~2 of the algorithm is equivalent to
\begin{align}
 \Expe{\rvey^{1}_{k-1}\rvax_{k}}
 =
 \bA^{k-1_{\!\lrcorner}}
 \Expe{\rvex^{1}_{k-1}\rvax_{k}}.
\end{align}
This means that the top~$(k-1)$ entries in the $k$-th column of $\bK_{\rvey^{1}_{k}\rvex^{1}_{k}}$ depend only on the entries of $\bK_{\rvex^{1}_{k}}$ above its $k$-th row.
Recalling that $\bK_{\rvey^{1}_{k}\rvex^{1}_{k}} = \bA\bK_{\rvex^{1}_{k}}$, we conclude that $\bA$ is also lower triangular, and thus $\rvey^{1}_{k}$ also satisfies causality.
Notice that for any given $\bK_{\rvex^{1}_{k-1}}$ and $\bK_{\rvey^{1}_{k-1}\rvex^{1}_{k-1}}$ satisfying causality up to sample $k-1$, the vector $\Expe{\rvey^{1}_{k-1}\rvax_{k}}$ yielded by Step~2 is the only vector consistent with $\rvax^{k},\rvay^{k}$ satisfying causality up to the $k$-th sample.
\paragraph*{MMSE Condition~\eqref{eq:z_perp_y}}
Step~1 guarantees that~\eqref{eq:z_perp_y} is satisfied for $k=1$.
Steps 3, 4 and 5 mean that $\Expe{\rvey^{1}_{k}\rvay_{k}}=\Expe{\rvey^{1}_{k}\rvax_{k}}$ for all $k=2,\ldots,\ell$.
Therefore, the reconstruction vector $\rvey^{1}_{\ell}$ yielded by the above algorithm satisfies~\eqref{eq:z_perp_y} for all $k=1,\ldots,\ell$.

\paragraph*{Source's past independence~\eqref{eq:MC_key}}
Since all variables are jointly Gaussian, condition~\eqref{eq:MC_key} is equivalent to 
\begin{align}\label{eq:Ey-Eyxy=0}
\Expe{(\rvay_{k} -\expe{\rvay_{k}|\rvax_{k},\rvey^{1}_{k-1}})(\rvex^{1}_{k-1})^{T}} =\bzero,
\end{align}
for all $k=1,\ldots, \ell$.
On the other hand,
\begin{align}\label{eq:lagrande}
 \Expe{\rvay_{k}|\rvax_{k},\rvey^{1}_{k-1}}
 &
 =
 \Expe{\rvay_{k}[(\rvey^{1}_{k-1})^{T} \, \rvax_{k}]}
 \left(
\begin{matrix}
 \bK_{\rvey^{1}_{k-1}} 					& \Expe{\rvax_{k}\rvey^{1}_{k-1}}\\
 \Expe{\rvax_{k}\rvey^{1}_{k-1}}^{T} 	& \Expe{\rvax_{k}^{2}}
\end{matrix}
\right)^{-1}
\left[
\begin{matrix}
 \rvey^{1}_{k-1}\\
 \rvax_{k}
\end{matrix}
\right].
\end{align}
From steps 1,~3 and~4 it follows that 
$
 \Expe{\rvay_{k}[(\rvey^{1}_{k-1})^{T} \, \rvax_{k}]} =
 \Expe{\rvay_{k}(\rvey^{1}_{k})^{T}}
 =
 \Expe{\rvax_{k}(\rvey^{1}_{k})^{T}}
$.
Substitution of this into~\eqref{eq:lagrande} and the result into~\eqref{eq:Ey-Eyxy=0} leads directly 
to~\eqref{eq:Past_indep_trans}.
Thus,~\eqref{eq:MC_key} is satisfied for all $k=1,\ldots,\ell$. 

Since the above algorithm yields an output which satisfies~\eqref{eq:MC_causality},~\eqref{eq:z_perp_y} and~\eqref{eq:MC_key}, for all $k=1,\ldots,\ell$, this output attains equality in~\eqref{eq:IgeqLB}, thus being a realization of $R_{\ell}^{SRD}(D_{1},\ldots,\D_{\ell})$.
Notice that once the distortions $\set{d_{k}}_{k=1}^{\ell}$ are given, each step in the recursive algorithm yields the only variances and covariances that satisfy~\eqref{eq:MC_causality},~\eqref{eq:z_perp_y} and~\eqref{eq:MC_key}. 
Therefore, for any given distortion schedule $\set{D_{k}}_{k=1}^{\ell}$, the latter algorithm yields the unique output that realizes $R_{\ell}^{SRD}(D_{1},\ldots, D_{\ell})$.
This completes the proof.
\findemo


\section{Proof of Theorem~\ref{thm:shortone}}\label{sec:proof_shortone}
Consider the first~$\ell$ samples of input and output.
The average distortion constraint here takes the form
\begin{align}\label{eq:2/ksumsigza}
 \frac{1}{\ell} \Sumfromto{k=1}{\ell}\sigsq_{\rvaz(k)} \leq D.
\end{align}
%
Then, 
%
\begin{align}
R_{c}^{it(\ell)}(D)
&\eq
\inf_{\rvay^{\ell}:
\text{\eqref{eq:MC_causality} and \eqref{eq:2/ksumsigza} hold}
}
\tfrac{1}{\ell} I(\rvax^{\ell};\rvay^{\ell})
=
\inf_{\set{\rvaz(k)}_{k=1}^{\ell}:
\eqref{eq:2/ksumsigza} \text{ holds}}
R_{\ell}^{SRD}(\sigsq_{\rvaz(1)},\ldots, \sigsq_{\rvaz(\ell)})\nonumber
\\&
= 
\inf_{\set{\rvaz(k)}_{k=1}^{\ell}:
\eqref{eq:2/ksumsigza} \text{ holds}}
\Set{
\frac{1}{2\ell}\ln\left(
\frac{\sigsq_{\rvax}}{\sigsq_{\rvaz(\ell)}} 
\right)
+
\frac{1}{2\ell}\sumfromto{k=1}{\ell-1}\ln\left( 
\frac{a^{2}\sigsq_{\rvaz(k)} +\sigsq_{\xi}}
{\sigsq_{\rvaz(k)}}
\right)}\nonumber
\\&
\geq
\inf_{\set{\rvaz(k)}_{k=1}^{\ell}:
\eqref{eq:2/ksumsigza} \text{ holds}}
\Set{
\frac{1}{2\ell}\ln\left( \frac{\sigsq_{\rvax}}{\sigsq_{\rvaz(\ell)}}\right)
+
\frac{(\ell-1)}{2\ell}\ln\left( 
a^{2}+\frac{\sigsq_{\xi}}
{\frac{1}{\ell-1}\sumfromto{k=1}{\ell-1}\sigsq_{\rvaz(k)}}
\right)}\label{eq:firstLB}
\end{align}
where the last inequality follows from Jensen's inequality and the fact that $\ln(a^{2}+\frac{b^{2}}{x})$ is a convex function of $x$.
Equality is achieved if and only if all distortions $\sigsq_{\rvaz(k)}$ equal some common value for all $k=1,\ldots,(\ell-1)$.
Given that the RHS of~\eqref{eq:firstLB} is minimized when constraint~\eqref{eq:2/ksumsigza} is active 
(i.e., by making $\frac{1}{\ell}\sumfromto{k=1}{\ell}\sigsq_{\rvaz(k)}=D$), we can attain equality in~\eqref{eq:firstLB} and minimize its RHS by picking
\begin{align}\label{eq:allequal}
\sigsq_{\rvaz(k)}= \frac{\ell D- \sigsq_{\rvaz(\ell)}}{\ell-1},\fspace \forall k\in\set{1,2,\ldots,\ell-1}.
\end{align}
For this choice to be feasible, the distortion $\sigsq_{\rvaz(k)}$ must satisfy~\eqref{eq:subeq_eff_sigmas}, which translates into the constraint
\begin{align}\label{eq:sigsqzell_leq_something}
\sigsq_{\rvaz(\ell)}
\leq 
\frac{\ell a^{2}D +(\ell-1)\sigsq_{\xi}}
{\ell-1 +a^{2}}
\eq 
U(\ell). 
\end{align}
Thus, substituting~\eqref{eq:allequal} into~\eqref{eq:firstLB}, we obtain
\begin{align}\label{eq:casi}
 R_{c}^{it(\ell)}(D)
=
\inf_{\rvaz(\ell):
\sigsq_{\rvaz(\ell)}\leq U(\ell)}
\Set{
\frac{1}{2\ell}\ln\left(\frac{\sigsq_{\rvax}}{\sigsq_{\rvaz(\ell)}} \right)
+
\frac{(\ell-1)}{2\ell}\ln\left( 
a^{2}+\frac{(\ell-1)\sigsq_{\xi}}
{\ell D -\sigsq_{\rvaz(\ell)}}
\right)
}.
\end{align}
In view of~\eqref{eq:sigsqzell_leq_something}, as $\ell\to\infty$, the 
value of $\sigsq_{\rvax(\ell)}$ that infimizes~\eqref{eq:casi} remains bounded.
Therefore,
\begin{align}\label{eq:ellmite}
 \lim_{\ell\to\infty} R_{c}^{it(\ell)}(D) 
 =
 \max\Set{0\,,\,
\frac{1}{2}\ln\left( 
a^{2}+\frac{\sigsq_{\xi}}
{ D }
\right)
}
\end{align}
Finally, from Lemma~\ref{lem:limusp} in the Appendix, we conclude that $R_{c}^{it}(D)$ equals the RHS of~\eqref{eq:ellmite}, completing the proof.
\findemo


\section{Proof of Theorem~\ref{thm:Bouds_B}}\label{sec:proof_Bouds_B}
The first inequality in~\eqref{eq:ineqs_B} follows directly from definitions~\ref{def:CausalRDF}
and~\ref{def:Stat_CausalRDF}.
For a plain AWGN channel with noise variance $d$, the mutual information between source and
reconstruction is
\begin{align*}
 R_{AWGN}(d) 
\eq \frac{1}{4\pi}
\intfromto{-\pi}{\pi}
\log_{2}\left(1+\frac{S_{\rvax}\ejw}{d}\right)d\w.
\end{align*}
On the other hand, by definition, the mutual information across a test channel that realizes
$R^{\perp}(D)$ with distortion $D=d$ satisfies~\cite{derost08}:
\begin{align*}
R^{\perp}(d)\leq R_{AWGN}(d).
\end{align*}
In both cases the end-to-end distortion can be reduced by placing a scalar gain after the test
channel.
The optimal (minimum MSE) gain is $\frac{\sigsq_{\rvax}}{\sigsq_{\rvax}+d}$. 
The mutual information from the source to the signal before the scalar gain is the same as that
between the source an the signal after it.
However, now the resulting end-to-end distortion is 
$D=\frac{d\sigsq_{\rvax}}{\sigsq_{\rvax}+d}$.
Therefore, for a given end-to-end distortion $D$, the distortion between the source and the signal
before the optimal scalar gain is
\begin{align*}
 d = \frac{\sigsq_{\rvax}D}{\sigsq_{\rvax} -D},
\end{align*}
which implies that the mutual informations across the $R^{\perp}$ channel and the AWGN channel when
the optimal scalar gain is used are given by
$R^{\perp}(\frac{\sigsq_{\rvax}D}{\sigsq_{\rvax} -D} )$ and 
$R_{AWGN}(\frac{\sigsq_{\rvax}D}{\sigsq_{\rvax} -D} )$, respectively.
We then have that
{\allowdisplaybreaks 
\begin{subequations}\label{eq:Boudsineqs}
\begin{align}
\overline{R_{c}^{it}}(D)
-R(D)
& 
\leq  R^{\perp}(\tfrac{\sigsq_{\rvax}D}{\sigsq_{\rvax} -D} ) - R(D)
=
B_{2}(D)
\nonumber
\\&
\leq R_{AWGN}(\tfrac{\sigsq_{\rvax}D}{\sigsq_{\rvax} -D}) - R(D)\nonumber
=
\frac{1}{4\pi}\intfromto{-\pi}{\pi}\log_{2}\left(1+\frac{S_{\rvax}\ejw}{\frac{\sigsq_{\rvax}D}{
\sigsq_{\rvax} -D}}\right)d\w
-
R(D)\nonumber
\\&
=
\frac{1}{4\pi}\intfromto{-\pi}{\pi}\log_{2}\left(1+[1-\tfrac{D}{\sigsq_{\rvax}}]\frac{S_{\rvax}\ejw}
{D} \right)d\w
-
R(D)
=
B_{2}(D).
\label{eq:aca}
\end{align}
\end{subequations}
}
%

To obtain the first function within the $\min$ operator on the RHS of~\eqref{eq:B4}, 
we notice from~\eqref{eq:R(D)waterfill} that, 
since $\vareps\leq D\leq \theta$, the RDF for a Gaussian stationary source with PSD $S_{\rvax}^{\vareps}\ejw\eq \max\Set{\vareps,S_{\rvax}\ejw}$, $\forall \w \in\pipi$, 
say $R^{\vareps}(\cdot)$, will equal the value $R(D)$ given by~\eqref{eq:R(D)waterfill_R(D)} when the ``water level'' $\theta$ takes the same value as in~\eqref{eq:R(D)waterfill}.
Hence, denoting by $D^{\vareps}$ the distortion obtained in~\eqref{eq:R(D)waterfill} when $S_{\rvax}$
is substituted by $S^{\vareps}_{\rvax}$, we find that
\begin{align}\label{eq:D<D+e}
 R^{\vareps}(D^{\vareps}) =R(D) \iff D^{\vareps} = 
 \frac{1}{2\pi}\Intfromto{-\pi}{\pi}\min\Set{\theta, S^{\vareps}_{\rvax}\ejw}d\w
 \leq 
 D+\vareps.
\end{align}

On the other hand,
\begin{align}\label{eq:RvarepsD<IntS/D}
 R^{\vareps}(D^{\vareps}) 
 \geq 
 \frac{1}{4\pi}\Intfromto{-\pi}{\pi}\log_{2}\left( \frac{S^{\vareps}\ejw}{D^{\vareps}}\right)d\w
\end{align}
With this, and starting from~\eqref{eq:aca}, we have the following:
\begin{align}
  \overline{R_{c}^{it}}(D)-R(D)
  &
  \leq
  \frac{1}{4\pi}\intfromto{-\pi}{\pi}\log_{2}\left(1+[1-\tfrac{D}{\sigsq_{\rvax}}]\frac{S_{\rvax}\ejw}{D} \right)d\w
  -R(D)\nonumber
  \\&
  \leq 
  \frac{1}{4\pi}\intfromto{-\pi}{\pi}\log_{2}\left(1+[1-\tfrac{D}{\sigsq_{\rvax}}]\frac{S^{\vareps}_{\rvax}\ejw}{D} \right)d\w
  -
  \frac{1}{4\pi}\Intfromto{-\pi}{\pi}\log_{2}\left( \frac{S^{\vareps}\ejw}{D^{\vareps}}\right)d\w 
  \label{eq:intlog-intlog}
  \\&
  =
   \frac{1}{4\pi}
   \intfromto{-\pi}{\pi}\log_{2}
   \left(
   \frac{D^{\vareps}}{S^{\vareps}\ejw}+[1-\tfrac{D}{\sigsq_{\rvax}}]\frac{D^{\vareps}}{D} 
   \right)d\w
   \\&
   \leq 
   \frac{1}{4\pi}
   \intfromto{-\pi}{\pi}\log_{2}
   \left(
   \frac{D+\vareps}{S^{\vareps}\ejw}+[1-\tfrac{D}{\sigsq_{\rvax}}]\frac{D+\vareps}{D}
   \right)d\w\label{eq:greater}
   \\&
   \leq 
   \frac{1}{2}
   \log_{2}
   \left(
   (D+\vareps)
  \varsigma_{\rvax}^{\vareps} +[1-\tfrac{D}{\sigsq_{\rvax}}]\frac{D+\vareps}{D}
   \right),\label{eq:38a}
  \end{align}
 where~\eqref{eq:intlog-intlog} follows from~\eqref{eq:R(D)waterfill},~\eqref{eq:D<D+e} and~\eqref{eq:RvarepsD<IntS/D} and by noting that
$S_{\rvax}^{\vareps}\ejw\geq S_{\rvax}\ejw,\,\forallwinpipi$,%
~\eqref{eq:greater} stems from~\eqref{eq:D<D+e}, and~\eqref{eq:38a} follows from Jensen's inequality. 
Notice that the RHS of~\eqref{eq:38a} equals the first term on the RHS of~\eqref{eq:B4}.
%

The middle term on the RHS of~\eqref{eq:B4} follows directly from~\eqref{eq:bounds_Rop_Rit_RD}.
Finally, for distortions close to $\sigsq_{\rvax}$, a bound tighter than~\eqref{eq:38a} can be obtained
from~\eqref{eq:aca} as follows
\begin{subequations}\label{eq:More_bound_ineqs}
\begin{align}
\overline{R_{c}^{it}}(D)-R(D)
& \leq  
B_{3}(D)
=
\frac{1}{4\pi}\intfromto{-\pi}{\pi}\log_{2}\left(1+\frac{[\sigsq_{\rvax}-D]S_{\rvax}\ejw}{\sigsq_{
\rvax}D}\right)d\w
 - R(D)
\nonumber
\\&
\label{eq:B4a}
<
\frac{1}{4\pi}\intfromto{-\pi}{\pi}\log_{2}\left(1+\frac{[\sigsq_{\rvax}-D]S_{\rvax}\ejw}{\sigsq_{
\rvax}D}\right)d\w
\\&
\label{eq:B4b}
\leq
\frac{1}{2}\log_{2}\left(1 + \frac{\sigsq_{\rvax} -D}{D} \right)
=
\frac{1}{2}\log_{2}\left(\frac{\sigsq_{\rvax}}{D} \right),
\end{align}
\end{subequations}
which is precisely the third term on the RHS of~\eqref{eq:B4}.
In the above,~\eqref{eq:B4a} holds trivially since $R(D)> 0,\, \forall D<\sigsq_{\rvax}$,
and~\eqref{eq:B4b} follows from Jensen's inequality.
Therefore, equality holds in~\eqref{eq:B4b} if and only if $\procx$ is white.
The validity of the chain of inequalities in~\eqref{eq:ineqs_B} follows directly
from~\eqref{eq:Boudsineqs} and~\eqref{eq:More_bound_ineqs}.
This completes the proof.
\findemo

\section{Proof of Lemma~\ref{lem:filters_realize_Rc}}\label{sec:proof_filters_realize_Rc}
The idea of the proof is to first show that if the distortion $D_{c}$ equals $D>0$, then
\begin{align}\label{eq:ineq_chain}
		\frac{1}{2}\ln(K)
		=
		I(\rvav(k);\rvaw(k))
		\overset{(a)}{\geq}
		\bar{I}(\procx;\procy)
		\overset{(b)}{\geq}
		\overline{R_{c}^{it}}(D).
	\end{align}

Immediately afterward we prove that, despite the distortion and causality constraints, the scheme 
in Fig.~\ref{fig:causalFQ} has enough degrees of freedom to turn all the above inequalities into equalities.
That means that if we are able to globally infimize $K$ over the filters of the system while satisfying the distortion and causality constraints, then that infimum, say $K_{inf}$, must satisfy $(1/2)\ln(K_{inf})
=
\overline{R_{c}^{it}}(D_{c})$.

We now proceed to demonstrate the validity of~\eqref{eq:ineq_chain} and to state the conditions under which equalities are achieved.
The first equality in~\eqref{eq:ineq_chain} follows from the fact that $\procn$ is a Gaussian i.i.d. process.
	Inequality~$(a)$ stems from the following:
{\allowdisplaybreaks
\begin{align}
I(\rvav(k); \rvaw(k)) 
&
= h(\rvaw(k)) - h(\rvaw(k)|\rvav(k))
= h(\rvaw(k)) - h(\rvav(k)+\rvan(k)|\rvav(k))\nonumber
\\&
= h(\rvaw(k)) - h(\rvan(k)|\rvav(k))\nonumber
\\&
= h(\rvaw(k)) - h(\rvan(k))\label{eq:n_indep_v}
\\&
\geq 
h(\rvaw(k)|\rvaw^{k-1}) - h(\rvan(k))\label{eq:w_iid}
\\&
= \bar{h}(\procw) - h(\rvan(k)|\rvan^{k-1})\label{eq:n_iid}
\\&
=
\bar{h}(\procw) -
h(\rvan(k)|\rvan^{k-1},\rvav^{k})\label{eq:v_indep}
\\&
=
\bar{h}(\procw) - h(\rvaw(k)| \rvaw^{k-1},\rvav^{k})\label{eq:w_for_n}
\\&
=
\bar{h}(\procw) - h(\rvaw(k)| \rvaw^{k-1}, \tilde{\rvax}^{k})\label{eq:v_for_xtd}
\\&
=
\bar{h}(\procw) - h(\rvaw(k)| \rvaw^{k-1}, \tilde{\rvax}^{\infty})
\label{eq:dir_to_notdir}
\\&
=
\bar{I}(\procxtd;\procw),\nonumber
\\&
\geq 
\bar{I}(\procx;\procy)\label{eq:ABdataprocineq}
\end{align}
}%
where $\procxtd$ is the signal at the output of $A(z)$, see Fig.~\ref{fig:causalFQ}.
In the above,~\eqref{eq:n_indep_v} follows from the fact that $\procn$ and $\procx$ are independent
and from the fact that $F(z)$ is strictly causal.
As a consequence, $\rvan(k)$ is independent of $\rvav(k)$, for all $k\in\Z^{+}$.
Inequality~\eqref{eq:w_iid} holds from the property $h(x|y)\leq h(x)$, with equality if and only if
$x$ and $y$ are independent, i.e., if and only if $\procw$ is white.
Similarly,~\eqref{eq:n_iid} holds since the samples of $\procn$ are independent.
By noting that $\rvav^{k}$ is a linear combination of $\rvax^{k}$ and $\rvan^{k-1}$, it follows
immediately that $\rvan(k)$ is independent from $\rvav^{k}$ upon knowledge of $\rvan^{k-1}$, which
leads to~\eqref{eq:v_indep}.
On the other hand,~\eqref{eq:w_for_n} stems from the fact that 
$\rvaw^{k}=\rvan^{k}+\rvav^{k}$.
Equality in~\eqref{eq:v_for_xtd} holds from the fact that, if $\rvaw^{k-1}$ is known, then
$\tilde{\rvax}^{k}$ can be obtained deterministically from $\rvav^{k-1}$, and vice-versa, see
Fig.~\ref{fig:causalFQ}.
Equality~\eqref{eq:dir_to_notdir} follows from the fact that there exists no feedback from $\procw$
to $\procxtd$, and thus the Markov chain
$
\tilde{\rvax}_{k+1}^{\infty} 
\leftrightarrow
(\tilde{\rvax}^{k},\rvaw^{k-1})
\leftrightarrow
\rvaw(k)
$
holds.
On the other hand, $\bar{I}(\procxtd;\procw) \geq  \bar{I}(\procx;\procy)$,  with equality if and
only if $B\ejw$ is invertible for all frequencies $\w$ for which $\abs{A\ejw}>0$.
Finally,~\eqref{eq:ABdataprocineq} follows directly from the Data Processing Inequality, with equality if and only if 
$B\ejw $ is invertible for all frequencies $\w$ for which
$\abs{A\ejw} > 0$.

Since $\overline{R_{c}^{it}}(D)$ is by definition an infimum, it follows that, for every $\vareps>0$, there exists an output process $\set{\rvay'(k)}$ jointly Gaussian with $\procx$, satisfying the causality and distortion constraints and such that 
$
\bar{I}(\procx;\set{\rvay'(k)})\leq \overline{R_{c}^{it}}(D)+\vareps
$.
Such output can be characterized by its noise PSD, say $S'_{u}$, and its signal transfer function, say $W'(z)$, by using the model in Fig.~\ref{fig:roles}.

Therefore, all that is needed for the system in Fig.~\ref{fig:causalFQ} 
to achieve 
\begin{align}\label{eq:la69}
 \frac{1}{2}\ln(K)=\bar{I}(\procx;\set{\rvay'(k)})\leq \overline{R_{c}^{it}}(D)+\vareps
\end{align}
is to yield the required noise PSD $S'_{u}$, the required signal transfer function $W'(z)$, a white $\procw$ and satisfy $B\ejw \neq 0,\, \forall w:A\ejw \neq 0$.
To summarize and to restate the latter more precisely:
\begin{subequations}\label{eq:allconditions}
 \begin{align}
\fspace\text{Equality in~(51) $ \Leftarrow$ }
&
S_{\rvaw}\ejw = 1= \abs{A\ejw}^{2}S_{\rvax}\ejw + \abs{1 - F\ejw}^{2}\sigsq_{\rvan}\label{eq:Sw}
\\
\fspace\text{Equality in (new)~\eqref{eq:ABdataprocineq} $\Leftrightarrow$}
&
 B\ejw \neq 0,\, \forall \w: {A\ejw} \neq 0 \label{eq:B=A-1}
\\
\fspace\text{\eqref{eq:la69} holds  $\Leftarrow$}
& 
\Biggl\{
\begin{array}{rl}
	W\ejw &= W'\ejw\\\
	S_{\rvau}'\ejw &=\abs{W'\ejw}^{2}\abs{B\ejw}^{2}\abs{1-F\ejw}^{2}\sigsq_{\rvan}
\end{array}
\label{eq:WandSu}
\end{align}
\end{subequations}
All these equations are to be satisfied $\aeonpipi$.
We have chosen $\sigsq_{\rvaw}=1$ in~\eqref{eq:Sw} for simplicity and because, as we shall see next, we have enough degrees of freedom to do so without compromising rate/distortion performance.
Solving the system of equations formed by~\eqref{eq:Sw},~\eqref{eq:WandSu} and~\eqref{eq:B=A-1} we obtain
\begin{subequations}\label{eq:lasexplicitas}
\begin{align}
 \abs{B\ejw}^{2}
 &=
 \frac{S'_{\rvau}\ejw +  \abs{W'\ejw}^{2}S_{\rvax}\ejw}{\abs{W'\ejw}^{2}}\label{eq:|B|^2} \fspace \aeonpipi
 \\
 \abs{1-F\ejw}^{2}\sigsq_{\rvan}
 &=
 \frac{S'_{\rvau}\ejw}{S'_{\rvau}\ejw +  \abs{W'\ejw}^{2}S_{\rvax}\ejw}\label{eq:|1-F|^2} \fspace \aeonpipi
\\
\abs{A\ejw}^{2}
&=
\frac{\abs{W\ejw}^{2}}
{S'_{\rvau}\ejw + \abs{W\ejw}^{2} S_{\rvax}\ejw}  
 \label{eq:|A|^2} \fspace \aeonpipi
\end{align}
\end{subequations}
It is only left to be shown that there exist causal, stable and minimum-phase transfer functions $B(z)$, $(1-F(z))$ and $A(z)$ such that their squared magnitudes equal their right-hand sides in~\eqref{eq:lasexplicitas}. 
To do so, we will make use of the Paley-Wiener theorem (Theorem~\ref{thm:Paley-Wiener} in the Appendix).

To begin with, we notice from Fig.~\ref{fig:roles}, and since $\set{\rvau'(k)}$ is independent of $\procx$, that 
\begin{align}
\bar{I}(\procx;\set{\rvay'(k)})
&=
\frac{1}{2}\Intfromto{-\pi}{\pi}
\ln\left(\frac{\abs{W'\ejw}^{2}S_{\rvax}\ejw +S'_{\rvau}}{S'_{\rvau}\ejw} \right)d\w
\\&=
\frac{1}{2}\Intfromto{-\pi}{\pi}
\abs{\ln\left(\abs{1-F\ejw}^{2}\sigsq_{\rvan} \right)}d\w\label{eq:rcit=int},
\end{align}
where~\eqref{eq:rcit=int} follows from~\eqref{eq:|1-F|^2}.
Since $\overline{R_{c}^{it}}(D)$ is bounded, so is $\bar{I}(\procx;\set{\rvay'(k)})$, and thus we conclude from the Paley-Wiener theorem that there exists a stable, causal and minimum-phase transfer function $(1-F(z))$ satisfying~\eqref{eq:|1-F|^2}.
Also, from the fact that the first sample of the impulse response of $(1-F(z))$ is~$1$ and as a consequence of $(1-F(z))$ being minimum-phase, we conclude that $\intfromto{-\pi}{\pi}\ln\abs{1-F\ejw}d\w=0$ (see, e.g.,~\cite{serbra97}).
Therefore,
\begin{align}
\sigsq_{\rvan} 
  =  \expo{2\bar{I}(\procx;\set{\rvay'(k)})}.
\end{align}

Next, we notice that since $W(z)$ is stable and causal, then there exists a causal, stable and minimum phase transfer function $\widetilde{W}(z)$ such that $\abs{\widetilde{W}\ejw}=\abs{W\ejw}$, forall $\w\in\pipi$.
From the Paley-Wiener theorem, it follows that
\begin{align}
 \Intfromto{-\pi}{\pi}\abs{\ln\abs{\widetilde{W}\ejw} }d\w <\infty,
\end{align}
which implies that
\begin{align}\label{eq:-inf+inf}
-\infty< \Intfromto{-\pi}{\pi} \ln\abs{\widetilde{W}\ejw }d\w = 
\Intfromto{-\pi}{\pi} \ln\abs{W\ejw}d\w<\infty.
\end{align}
On the other hand, from~\eqref{eq:rcit=int}, 
\begin{align}
 \overline{R_{c}^{it}}(D) &
 \geq
 \frac{1}{2}\Intfromto{-\pi}{\pi}
\ln\left(\frac{\abs{W'\ejw}^{2}S_{\rvax}\ejw}{S'_{\rvau}\ejw} \right)d\w
\end{align}
and recalling that 
$
\abs{\intpipi{\ln S_{\rvax}\ejw}}< \infty
$,
it follows that  $\intfromto{-\pi}{\pi}\ln(S'_{\rvau}\ejw/\abs{W\ejw}^{2})d\w$ is bounded from below.
In view of~\eqref{eq:-inf+inf}, we conclude that 
$\intfromto{-\pi}{\pi}\ln(S'_{\rvau}\ejw)d\w>-\infty$.
Now, since $\frac{1}{2\pi}\intfromto{-\pi}{\pi}S'_{\rvau}\ejw d\w \leq D$, we can apply Lemma~\ref{lem:paraPaleyWiener} (see Appendix) to obtain that
\begin{align}
 \Intfromto{-\pi}{\pi}\abs{\ln S'_{\rvau}\ejw}d\w.
 < \infty
\end{align}
Substitution of the RHS of the second equation of~\eqref{eq:WandSu} into the above, together with the Paley-Wiener theorem, yields that there exists a causal, stable and minimum phase transfer function $G(z)$ such that
\begin{align}
 \abs{G\ejw}^{2} = \abs{\widetilde{W}\ejw}^{2}\abs{B\ejw}^{2}\abs{1-F\ejw}^{2}\sigsq_{\rvan},
\end{align}
and thus $B(z)$ can be chosen to be the causal, stable and minimum-phase transfer function
\begin{align}
 B(z)=
\frac{G(z)}{\widetilde{W}(z) (1-F(z))\sigma_{\rvan}}.
\end{align}
which allows us to choose a stable, causal and minimum-phase $A(z)=B(z)^{-1}$.
Therefore, for every $\vareps>0$, there exists causal, stable and minimum phase transfer functions $A(z)$, $B(z)$ and $1-F(z)$ that satisfy~\eqref{eq:allconditions}, attaining equalities throughout and therefore yielding a value of $K$ which satisfies~\eqref{eq:la69}.  
This completes the proof.
\findemo


\section{Proof of Lemma~\ref{lem:Jsc()_is_convex}}\label{sec:proof_Jsc()_is_convex}
Strict convexity exists if and only if the inequality
\begin{align}
\lambda\Jsc(p_{1}) + [1-\lambda]\Jsc(p_{2}) >  \Jsc(\lambda p_{1}+[1-\lambda] p_{2}), \fspace\forall
\lambda\in(0,1),\label{eq:Jp1_Jp2_convexity}
\end{align}
holds for any two pairs $p_{1}\eq(f_1,g_1)\in\Fset_{K}\times\Gset$ and
$p_{2}\eq(f_2,g_2)\in\Fset_{K}\times\Gset$
satisfying
\begin{align}\label{eq:cond_fg_fg>0}
 \norm{f_{1}-f_{2}} + \norm{g_{1}-g_{2}}>0.
\end{align}
We will first prove the validity of~\eqref{eq:Jp1_Jp2_convexity} for pairs $p_{1}$ and $p_{2}$ which
also satisfy
\begin{align}\label{eq:g_0_>0}
\abs{\lambda g_{1}(\w) + [1-\lambda]g_{2}(\w) } >0,\,\forallwinpipi, \forall \lambda\in[0,1],
\end{align}
but are otherwise arbitrary.
For any given $\lambda\in[0,1]$, define the pair
\begin{equation*}
	(f_0 , g_0) \eq \lambda (f_1,g_1) + [1-\lambda] (f_2,g_2).
\end{equation*} 
Upon defining the functions
\begin{equation}\label{eq:eta_theta_def}
	\eta \eq f_2 - f_1;\fspace \theta \eq g_2 - g_1,
\end{equation} 
any pair along the ``line'' between $(f_1,g_1)$ and $(f_2,g_2)$ can be written in terms of a
single scalar parameter $s$ via
\begin{equation*}
	(f,g) = (f_0 + \eta s \, , \, g_0 + \theta s),
\end{equation*} 
where $s\in[\lambda-1,\lambda]$.
Define the functions
\begin{subequations}\label{eq:Nsp_and_Dsp}
 \begin{align}
 \Nsp(s) &\eq \ip{f,\abs{g}} = \Ip{f_{0}+\eta s , \hsqrt{\abs{g_{0}}^{2}  +
2\Rsp\set{g_{0}\theta^{*}} s + \abs{\theta}^{2} s^{2} } },\label{eq:Nsp_def}\\
 \Dsp(s) &\eq K-\norm{f}^{2} = K- \norm{f_{0}}^{2}- 2\ip{f_{0},\eta} s -
\norm{\eta}^{2}s^{2},\label{eq:Dsp_def}
\end{align}
\end{subequations}
where $\Rsp\set{x}$ denotes the real part of $x$.
Substitution of~\eqref{eq:Nsp_and_Dsp} into~\eqref{eq:V2} allows one to write the latter as
\begin{equation*}
\Jsc(f,g) =  
	J(s)\eq \frac{\Nsp(s)^{2}}{\Dsp(s)}
		+ L + a s + \norm{\theta}^{2} s^{2}
\end{equation*} 
where
\begin{eqnarray*}
a 		&\eq & 2 \,\Rsp\set{\ip{ g_0\!-\! G, \, \theta}} \\
L		&\eq & \norm{g_0}^{2} +\norm{G}^{2}-2 \Rsp\Set{\ip{g_0,G}}.
\end{eqnarray*} 
We next show that~\eqref{eq:Jp1_Jp2_convexity} holds by showing that $d^{2} J(s)/d s^{2}|_{s=0} >0$
for every $\lambda\in[0,1]$.
For this purpose, we first take the derivative of $J(s)$ with respect to $s$.
Denoting the derivatives of the functions $\Dsp(s)$ and $\Nsp(s)$ with respect to $s$ by $\Dsp'$ and
$\Nsp'$, respectively, we have that
\begin{align*}
 J'(s) 
=
\frac{2\Nsp \Nsp'\Dsp  - \Nsp^{2}\Dsp'}{\Dsp^{2}} + a + 2\norm{\theta}^{2}s.
\end{align*}
Differentiating again, one arrives to
\begin{align}
 J''(s) 
&=
\frac{
2\left(
\Nsp'\Dsp
-
\Nsp\Dsp'
\right)^{2}
+
2\Nsp\Nsp''\Dsp^{2}
-
\Nsp^{2}\Dsp''\Dsp
}
{\Dsp^{3}}
+2\norm{\theta}^{2}\nonumber\\
&=
\frac{
2\left(
\Nsp'\Dsp
-
\Nsp\Dsp'
\right)^{2}
+
(
2\Nsp\Nsp''\Dsp
-
\Nsp^{2}\Dsp''
+2\norm{\theta}^{2}\Dsp^{2}
)\Dsp
}
{\Dsp^{3}}.
\label{eq:J''(s)_last}
\end{align}
%
From~\eqref{eq:J''(s)_last}, we have that
\begin{align}\label{eq:J''geq}
 J''(s)_{|s=0}
&
=
\frac{2(\Nsp_{0}'\,\Dsp_{0}-\Nsp_{0}\Dsp_{0}')^{2}}{\Dsp_{0}^{3}}
+
\frac{2\Nsp_{0}\Nsp''_{0}\Dsp_{0}
-\Nsp_{0}^{2}\Dsp_{0}''+2\norm{\theta}^{2}\Dsp_{0}^{2}}{\Dsp_{0}^{2}}
\end{align}
where
\begin{equation}\label{eq:Nsp0_and_Dsp0}
\begin{split}
 \Nsp_{0}&\eq \Nsp(s)_{|s=0}=\ip{f_{0},\abs{g_{0}}}\\
 \Nsp'_{0}&\eq \Nsp(s)'_{|s=0} = \ip{f_{0},\frac{c}{\abs{g_{0}}}}\\
 \Nsp''_{0}&\eq \Nsp(s)''_{|s=0} = \Ip{f_{0}, \frac{\abs{\theta}^{2}\abs{g_{0}}^{2}- c^{2}  
}{\abs{g_{0}}^{3}} }+2\ip{\eta , \frac{c}{\abs{g_{0}}}}\\
\Dsp_{0} & \eq \Dsp(s)_{|s=0} = K - \norm{f_{0}}^{2}\\
\Dsp'_{0}  &\eq	\frac{\partial D}{\partial s}\Big\vert_{s=0} = -2\ip{f_0,\eta}\\
\Dsp''_{0}&\eq \Dsp(s)''_{|s=0} = -2\norm{\eta}^{2},
\end{split} 
\end{equation}
see~\eqref{eq:Nsp_and_Dsp},
and where
\begin{align}\label{eq:c_Rsp_def}
  c\eq \Rsp\set{g_{0}\theta^{*}}.
\end{align}
Notice that~$\Nsp_{0}'$ and $\Nsp_{0}''$ in~\eqref{eq:Nsp0_and_Dsp0} are well defined since we are
considering pairs $p_{1}$ and $p_{2}$ for which~\eqref{eq:g_0_>0} holds.

Substitution of~\eqref{eq:Nsp0_and_Dsp0} into~\eqref{eq:J''geq} yields
\begin{align}
 J''(s)_{|s=0}
&
\overset{\hphantom{(a)}}{=}
\frac{2(\Nsp_{0}'\,\Dsp_{0}-\Nsp_{0}\Dsp_{0}')^{2}}{\Dsp_{0}^{3}}
+
\frac{2\Nsp_{0}\Dsp_{0}
\Ip{f_{0}, \frac{\abs{\theta}^{2}\abs{g_{0}^{2}}- c^{2}}{\abs{g_{0}}^{3}} }
+
4\Nsp_{0}\Dsp_{0}
\ip{\eta , \frac{c}{\abs{g_{0}}}}
+
2\Nsp_{0}^{2}\norm{\eta}^{2}
+
2\norm{\theta}^{2}\Dsp_{0}^{2}}
{\Dsp_{0}^{2}}
\nonumber
\\
&
\overset{(a)}{\geq}
\frac{2(\ip{f_{0},\frac{c}{\abs{g_{0}}} }\Dsp_{0} +2 \Nsp_{0}\ip{f_{0},\eta})^{2}}{\Dsp_{0}^{3}}
+
\frac{2\Nsp_{0}^{2}\norm{\eta}^{2}
+
4\Nsp_{0}\Dsp_{0}
\ip{\eta , \frac{c}{\abs{g_{0}}}}
+
2\norm{\theta}^{2}\Dsp_{0}^{2}}
{\Dsp_{0}^{2}}
\nonumber
\\
&
\overset{(b)}{=}
\frac{2\Ip{f_{0}\,,\,\Dsp_{0} \Rsp\set{\frac{g_{0}}{\abs{g_{0}} }\theta^{*}} + 2\Nsp_{0}\eta       
}^{2}      }{\Dsp_{0}^{3}}
+
\frac{2\Nsp_{0}^{2}\norm{\eta}^{2}
+
4\Nsp_{0}\Dsp_{0}
\Rsp\set{\ip{\eta , \frac{g_{0}}{\abs{g_{0}}}\theta^{*} }}
+
2\Norm{\theta}^{2}\Dsp_{0}^{2}}
{\Dsp_{0}^{2}}
\nonumber
\\&
\overset{\hphantom{(a)}}{=}
\frac{2 \left(\Rsp\Set{\Ip{f_{0}\,,\,  2\Nsp_{0}\eta +\Dsp_{0}\frac{g_{0}}{\abs{g_{0}}}\theta^{*}   
    }}\right)^{2}      }{\Dsp_{0}^{3}}
+
\frac{2
\norm{\Nsp_{0} \eta\ + \Dsp_{0} \frac{g_{0}}{\abs{g_{0}}}\theta^{*}}^{2}}
{\Dsp_{0}^{2}}
> 0\label{eq:term_ineq},
\end{align}
where
$(a)$ and
$(b)$ follow from~\eqref{eq:Nsp0_and_Dsp0},~\eqref{eq:c_Rsp_def} and from the fact that
$\Rsp\set{g_{0}\theta^{*} }\leq \abs{g_{0}}\abs{\theta}$.
The strict inequality in~\eqref{eq:term_ineq} stems from the fact that
$\norm{\eta}+\norm{\theta}>0$.
The latter follows directly from~\eqref{eq:eta_theta_def} and~\eqref{eq:cond_fg_fg>0}.
Therefore~\eqref{eq:Jp1_Jp2_convexity} holds for any two pairs $p_{1}=(f_{1},g_{1}),\,
p_{2}=(f_{2},g_{2}) \in\Fset_{K}\times\Gset$ satisfying~\eqref{eq:g_0_>0}.

We will show now that~\eqref{eq:Jp1_Jp2_convexity} also holds for pairs 
$p_{1}, p_{2}$ which do not satisfy~\eqref{eq:g_0_>0}.
The idea is to construct another pair, say $p_{1}^{\delta}$, $p_{2}^{\delta}$, ``close'' to $p_{1}$, $p_{2}$ and meeting~\eqref{eq:g_0_>0}, and then show that strict convexity along the straight line between $p_{1}^{\delta}$ and $p_{2}^{\delta}$ implies strict convexity along the straight line between $p_{1}$ and $p_{2}$.

For this purpose, define, for any given pairs $p_{1}= (f_{1},g_{1})\in\Fset_{K}\times\Gset$,
$p_{2}=(f_{2},g_{2})\in\Fset_{K}\times\Gset$, the family of functions
\begin{align*}
 h_{\delta}(\w) \eq 
\begin{cases}
\delta &, \textrm{ if } \abs{g_{1}(\w)} + \abs{g_{2}(\w)} = 0\\
\delta \hsqrt{-1}\frac{g_{1}(\w)}{\abs{g_{1}(\w)}} &, \textrm{ if }  \lambda g_{1}(\w)+
[1-\lambda]g_{2}(\w) =0 \textrm{ for some } \lambda\in(0,1),\\
0		&, \textrm{ in any other case}.
\end{cases}
\end{align*}
where $\delta>0$ is a scalar parameter.
The functions $h_{\delta}$ defined above exhibit the property (to be exploited below) that
\begin{align}\label{eq:g+h>0}
 \abs{\lambda \big[g_{1}(\w)+h_{\delta}(\w)\big]+ [1-\lambda] \big[g_{2}(\w) + h_{\delta}(\w) \big]}
> 0, \fspace \forall g_{1},g_{2}\in\Gset, \; \forall \delta>0,\,\forall \lambda\in(0,1).
\end{align}
Upon introducing the notation $p^{\delta}\eq p + (0,h_{\delta})$ and $g^{\delta}\eq g + h_{\delta}$,
it follows directly from~\eqref{eq:g+h>0} that $p_{1}^{\delta},\,p_{2}^{\delta}$
satisfy~\eqref{eq:g_0_>0} for \emph{all} pairs  $p_{1},p_{2}\in\Fset_{K}\times \Gset$.
Notice also that
\begin{align}\label{eq:norm_g-g^delta<delta}
 \norm{g - g^{\delta}} \leq \delta.
\end{align}
On the other hand, it is easy to show that $\Jsc(p)$ is uniformly continuous at $\lambda p_{1}+
[1-\lambda]p_{2}$ for any pairs $p_{1},p_{2}\in\Fset_{K}\times \Gset$ and for all
$\lambda\in[0,1]$.
In view of~\eqref{eq:norm_g-g^delta<delta}, uniform continuity of $\Jsc(p)$ means that, for every
$\vareps>0$, there exists $\delta=\delta(\vareps)>0$ such that
\begin{align}\label{eq:J(p^delta)-J(p)<e}
 \abs{\Jsc(p^{\delta}) -\Jsc(p)}<\vareps, \fspace\forall p=\lambda p_{1} + [1-\lambda]p_{2}, \,\forall \lambda\in(0,1).
\end{align}
The fact that $p_{1}^{\delta}$ and $p_{2}^{\delta}$ satisfy~\eqref{eq:g_0_>0} implies that
$p_{1}^{\delta}$, $p_{2}^{\delta}$ also satisfy the strict-convexity condition~\eqref{eq:Jp1_Jp2_convexity}.
Therefore, for each $\lambda\in(0,1)$, there exists $\vareps_{2}(\lambda)>0$ such that 
\begin{align}\label{eq:Jpdelta_e2}
\lambda \Jsc(p_{1}^{\delta})+[1-\lambda]\Jsc(p_{2}^{\delta}) -  \Jsc(\lambda p_{1}^{\delta}
+[1-\lambda]p_{2}^{\delta}) >\vareps_{2}(\lambda)>0 , \fspace \forall \lambda\in(0,1).
\end{align}
Then, from~\eqref{eq:J(p^delta)-J(p)<e} and~\eqref{eq:Jpdelta_e2},
\begin{align*}
 \lambda \Jsc(p_{1}) + [1-\lambda]\Jsc(p_{2})
&\geq
\lambda \Jsc(p_{1}^{\delta}) + [1-\lambda]\Jsc(p_{2}^{\delta}) - 2\vareps
\geq 
\Jsc(\lambda p_{1}^{\delta} +[1-\lambda]p_{2}^{\delta}) + \vareps_{2}(\lambda) - 2\vareps
\\&
\geq 
\Jsc(\lambda p_{1} +[1-\lambda]p_{2}) + \vareps_{2}(\lambda) - 3\vareps.
\end{align*}
Since $\delta$ can be chosen arbitrarily small, and in particular, strictly smaller than
$\delta(\vareps_{2}(\lambda)/3)>0$, it follows that~\eqref{eq:Jp1_Jp2_convexity} also holds for all pairs
$p_{1},\,p_{2}\in\Fset_{K}\times\Gset$ not satisfying~\eqref{eq:g_0_>0}.
This completes the proof.
\findemo


\section{Appendix}\label{sec:appendix}
\begin{lem}\label{lem:inflingeqliminf}
For any zero-mean Gaussian stationary source $\procx$ and $D>0$,
\begin{align}\label{eq:First_Part}
R_{c}^{it}(D) \geq \limsup_{k\to\infty}R_{c}^{it(k)}(D).
\end{align}
\end{lem}
\begin{proof}
Suppose~\eqref{eq:First_Part} does not hold, i.e., that
 \begin{align}\label{eq:limusps1}
  V
  \eq 
  \limsup_{k\to\infty} R_{c}^{it(k)}(D)
  =
  R_{c}^{it}(D)
  + \vareps_{1},
 \end{align}
for some $\vareps_{1}>0$.
The definition of $R_{c}^{it}(D)$ in~\eqref{eq:Rcitdef2} means that, $\forall \vareps_{2} >0$, there exists $\bar{\rvay}\in\Ssp$ such that
\begin{align}
    \limsup_{k\to\infty} \bar{I}(\rvax^{k};\bar{\rvay}^{k}) \leq  R_{c}^{it}(D) +\vareps_{2}
\end{align}
Combining this inequality with~\eqref{eq:limusps1} we arrive to
\begin{align}\label{eq:V2_V1_vareps}
 V=\limsup_{k\to\infty} \inf_{\rvay\in\Ssp} \bar{I}(\rvax^{k};\rvay^{k})
 \leq 
 \limsup_{k\to\infty} \bar{I}(\rvax^{k};\bar{\rvay}^{k})
 \leq 
 R_{c}^{it}(D)+\vareps_{2}
\end{align}

Since $\vareps_{2}$ can be chosen to be arbitrarily small, it can always be chosen so that $\vareps_{2} < \vareps_{1}$, which contradicts~\eqref{eq:limusps1}.
Therefore~\eqref{eq:First_Part} holds.
\end{proof}
 \begin{lem}\label{lem:limusp}
  Let 
  \begin{align}\label{eq:Rcitdef2}
  R_{c}^{it}(D)\eq \inf_{\procy\in\Ssp} \limsup_{k\to\infty}\bar{I}(\rvax^{k};\rvay^{k}),     
  \end{align}
  where $\Ssp$ denotes the space of all random processes causally related to $\procx$.
Let
\begin{align}
 R_{c}^{it(k)}(D)\eq \inf_{\rvay^{k}:\procy\in\Ssp}\bar{I}(\rvax^{k};\rvay^{k}).
\end{align}
Then, for any first-order Gauss-Markov source, the following holds:
\begin{align}
 R_{c}^{it}(D) =\limsup_{k\to\infty}R_{c}^{it(k)}(D).
\end{align}
\finenunciado
 \end{lem}

\begin{proof}
In Lemma~\ref{lem:inflingeqliminf} in the Appendix it is shown that
\begin{align}
R_{c}^{it}(D) \geq \limsup_{k\to\infty}R_{c}^{it(k)}(D),
\end{align}
so all we need to demonstrate is that 
$R_{c}^{it}(D) \leq \limsup_{k\to\infty}R_{c}^{it(k)}(D)$.
To do this, we simply observe from Theorem~\ref{thm:longone} that if we construct an output process $\procy$ by using the recursive algorithm of that theorem, with the choice $d_{k}=D$, for all $k\in\Nl$, then this output process is such that $\bar{I}(\procx;\procy)$ equals $V\eq\lim_{\ell\to\infty}R_{c}^{it(\ell)}(D)$.
Therefore, $R_{c}^{it}(D)\leq V$, concluding the proof.
\end{proof}

\begin{prop}[MMSE Column Correspondence]\label{prop:anyfiltercanbeWiener}
Let $\rvex\in\Rl^{k}$ be a Gaussian random  vector source with covariance matrix $\bK_{\rvex}$.
A reconstruction Gaussian random vector $\rvey$ satisfies
\begin{align}
 \Expe{\rvax_{k}|\rvey^{1}_{k}}=\rvay_{k}
\end{align}
if and only if
\begin{align}
 \bK_{\rvey}\be_{k,k} = \bK_{\rvey\rvex}\be_{k,k}.
\end{align}
\finenunciado
\end{prop}
\begin{proof}
We have that 
\begin{align}
  \bK_{\rvey\rvex}\be_{k,k}-\bK_{\rvey}\be_{k,k}
  =
  \Expe{\rvax_{k}\rvey^{1}_{k}} - \Expe{\rvay_{k}\rvey^{1}_{k}}
  =
  \Expe{(\rvax_{k}-\rvay_{k})\rvey^{1}_{k}}
\end{align}
The proof is completed by noting that $\Expe{\rvax_{k}|\rvey^{1}_{k}}=\rvay_{k}$ if and only if 
$ \Expe{(\rvax_{k}-\rvay_{k})\rvey^{1}_{k}}
  =0$.
\end{proof}

\begin{lem}[MMSE Triangular Correspondence]\label{lem:TR=TR}
Let $\rvex\in\Rl^{N}$, with $N\in\Nl$, be a Gaussian random source vector with covariance matrix $\bK_{\rvex}$.
A reconstruction Gaussian random vector $\rvey$ satisfies
\begin{align}\label{eq:Wiener_condition}
 \Expe{\rvax_{k}|\rvey^{1}_{k}}=\rvay_{k}, \fspace \forall k=1,2,\ldots N
\end{align}
if and only if
\begin{align}\label{eq:TR=TR}
 \left[\bK_{\rvey}\right]_{j,k} =\left[\bK_{\rvey\rvex}\right]_{j,k},\fspace \forall j\leq k, \;j,k=1,2,\ldots N.
\end{align}
\finenunciado
 \end{lem}
\begin{proof}
Let us first introduce the notation
$\bM^{k_{\!\lrcorner}}\in\Rl^{k\times k}$, denoting the top-left submatrix of any given square matrix $\bM\in{\Rl^{N\times N}}$, with $N\geq k$.
From Proposition~\ref{prop:anyfiltercanbeWiener}, it immediately follows that, for every $k=1,2,\ldots N$, 
\begin{align}
\bK_{\rvey}^{k_{\!\lrcorner}}
\be_{k,k}
=
\bK_{\rvey^{1}_{k}}
\be_{k,k}
=
\bK_{\rvey^{1}_{k}\rvex^{1}_{k}}
\be_{k,k}
=
\bK_{\rvey\rvex}^{k_{\!\lrcorner}}
\be_{k,k},
\end{align}
which is equivalent to~\eqref{eq:TR=TR}.
\end{proof}
Lemma~\ref{lem:TR=TR} implies that, if the reconstruction $\rvey$ is the output of a causal Wiener filter applied to the noisy source $\rvex+\rven$ for some noise vector $\rven$ (a condition equivalent to~\eqref{eq:Wiener_condition}), then $\bK_{\rvey}$ and $\bK_{\rvey\rvex}$ have identical entries on and above their main diagonals.

\paragraph*{Paley-Wiener Theorem}
\begin{thm}[{From~\cite[p.~229]{papoul77}} ]\label{thm:Paley-Wiener}
 Let $g\ejw$ be a non-negative function defined on $(-\pi,\pi]$.
 There exists a unique stable, causal and minimum phase transfer function $Y(z)$ such that $\abs{Y\ejw}^{2}=g\ejw$ if and only if%
 \footnote{In~\cite[p.~229]{papoul77} it is stated that~\eqref{eq:inabsg<infty} is a sufficient condition for such a $Y(z)$ to exist. 
 However, from~\cite[Note~2, p.~228]{papoul77} and the discrete-continuous equivalence in~\cite[p.~229]{papoul77} it follow that~\eqref{eq:inabsg<infty} is also necessary.}
 \begin{align}\label{eq:inabsg<infty}
  \Intfromto{-\pi}{\pi} \abs{ \log(g\ejw)}d\w <\infty
 \end{align}
\finenunciado
\end{thm}

\begin{lem}\label{lem:paraPaleyWiener}
 If $f(\w)\geq 0\forall\w\in\pipi$ and is such that 
 $\intfromto{-\pi}{\pi}f(\w)d\w <\infty$ 
 and 
 $\intfromto{-\pi}{\pi}\ln f(\w)d\w >-\infty$,
 then
 \begin{align}
  \Intfromto{-\pi}{\pi}\abs{\ln f(\w)}d\w <\infty
 \end{align}
\finenunciado
\end{lem}
\begin{proof}
Let $\Ssp\eq \set{\w\in\pipi : f(\w)\geq 1}$.
From Jensen's inequality and the fact that $\intfromto{-\pi}{\pi}f(\w)d\w <\infty$ , 
we have 
\begin{align}
 \Intover{\w\in \Ssp}\ln f(\w)d\w
 \leq 
 \abs{\Ssp}
 \ln\left(\frac{1}{\abs{\Ssp}} \intfromto{-\pi}{\pi}f(\w)d\w\right)
 <\infty.
\end{align}
This, together with
the condition  $\intfromto{-\pi}{\pi}\ln f(\w)d\w >-\infty$, implies that 
\begin{align}\label{eq:bouondedinf}
 -\Intover{\w\notin \Ssp} \ln f(\w)d\w <\infty
\end{align}
Therefore,
 \begin{align}
  \Intfromto{-\pi}{\pi}\abs{\ln f(\w)}d\w
  &=
  - 
  \Intover{\w\notin\Ssp} \ln f(\w)d\w
  +
  \Intover{\w\in\Ssp} \ln f(\w)d\w 
  <\infty ,
 \end{align}
completing the proof.
\end{proof}


\end{document}
    \fi